\documentclass{aastex62}

\graphicspath{{./}{figures/}}

\received{}
\revised{}
\accepted{}
\submitjournal{ApJ}

\shorttitle{Automatic detection of ICMEs}
\shortauthors{Nguyen et al.}

\begin{document}

\title{Automatic detection of Interplanetary Coronal Mass Ejections from in-situ data: a deep learning approach}

\correspondingauthor{Gautier Nguyen}
\email{gautier.nguyen@lpp.polytechnique.fr}

\author{Gautier Nguyen}
\affil{CNRS, Ecole polytechnique, Sorbonne Universit\'e, Univ Paris Sud, Observatoire de Paris, Universit\'e Paris-Saclay, PSL Research Univsersity, Laboratoire de Physique des Plasmas, Palaiseau, France}

\author{Nicolas Aunai}
\affiliation{CNRS, Ecole polytechnique, Sorbonne Universit\'e, Univ Paris Sud, Observatoire de Paris, Universit\'e Paris-Saclay, PSL Research Univsersity, Laboratoire de Physique des Plasmas, Palaiseau, France}

\author{Dominique Fontaine}
\affiliation{CNRS, Ecole polytechnique, Sorbonne Universit\'e, Univ Paris Sud, Observatoire de Paris, Universit\'e Paris-Saclay, PSL Research Univsersity, Laboratoire de Physique des Plasmas, Palaiseau, France}

\author{Erwan Le Pennec}
\affiliation{Ecole polytechnique, Centre de Mathématiques Appliqu\'ees, Palaiseau, France}

\author{Joris Van den Bossche}
\affiliation{Paris-Saclay Center for Data Science, INRIA, Palaiseau, France}

\author{Alexis Jeandet}
\affiliation{CNRS, Ecole polytechnique, Sorbonne Universit\'e, Univ Paris Sud, Observatoire de Paris, Universit\'e Paris-Saclay, PSL Research Univsersity, Laboratoire de Physique des Plasmas, Palaiseau, France}

\author{Brice Bakkali }
\affiliation{Ecole polytechnique, Palaiseau, France}

\author{Louis Vignoli}
\affiliation{Ecole polytechnique, Palaiseau, France}

\author{Bruno Regaldo-Saint Blancard}
\affiliation{Ecole polytechnique, Palaiseau, France}



\begin{abstract}
Decades of studies have suggested several criteria to detect Interplanetary coronal mass ejections (ICME) in time series from in-situ spacecraft measurements. Among them the most common are an enhanced and smoothly rotating magnetic field, a low proton temperature and a low plasma beta. However, these features are not all observed for each ICME due to their strong variability. Visual detection is time-consuming and biased by the observer interpretation leading to non exhaustive, subjective and thus hardly reproducible catalogs. Using convolutional neural networks on sliding windows and peak detection, we provide a fast, automatic and multi-scale detection of ICMEs. The method has been tested on the in-situ data from WIND between 1997 and 2015 and on the 657 ICMEs that were recorded during this period. The method offers an unambiguous visual proxy of ICMEs that gives an interpretation of the data similar to what an expert observer would give. We found at a maximum 197 of the 232 ICMEs of the 2010-2015 period (recall $84\pm 4.5 \%$) including $90\%$ of the ICMEs present in the lists of \citet{chinchilla} and \citet{Chi16}. The minimal number of False Positives was 25 out of 158 predicted ICMEs (precision $84 \pm 2.6 \%$).  Although less accurate, the method also works with one or several missing input parameters. The method has the advantage of improving its performance by just increasing the amount of input data. The generality of the method paves the way for automatic detection of many different event signatures in spacecraft in-situ measurements.

\end{abstract}



\section{Introduction}

\begin{figure}
\centering
\includegraphics[scale=0.9]{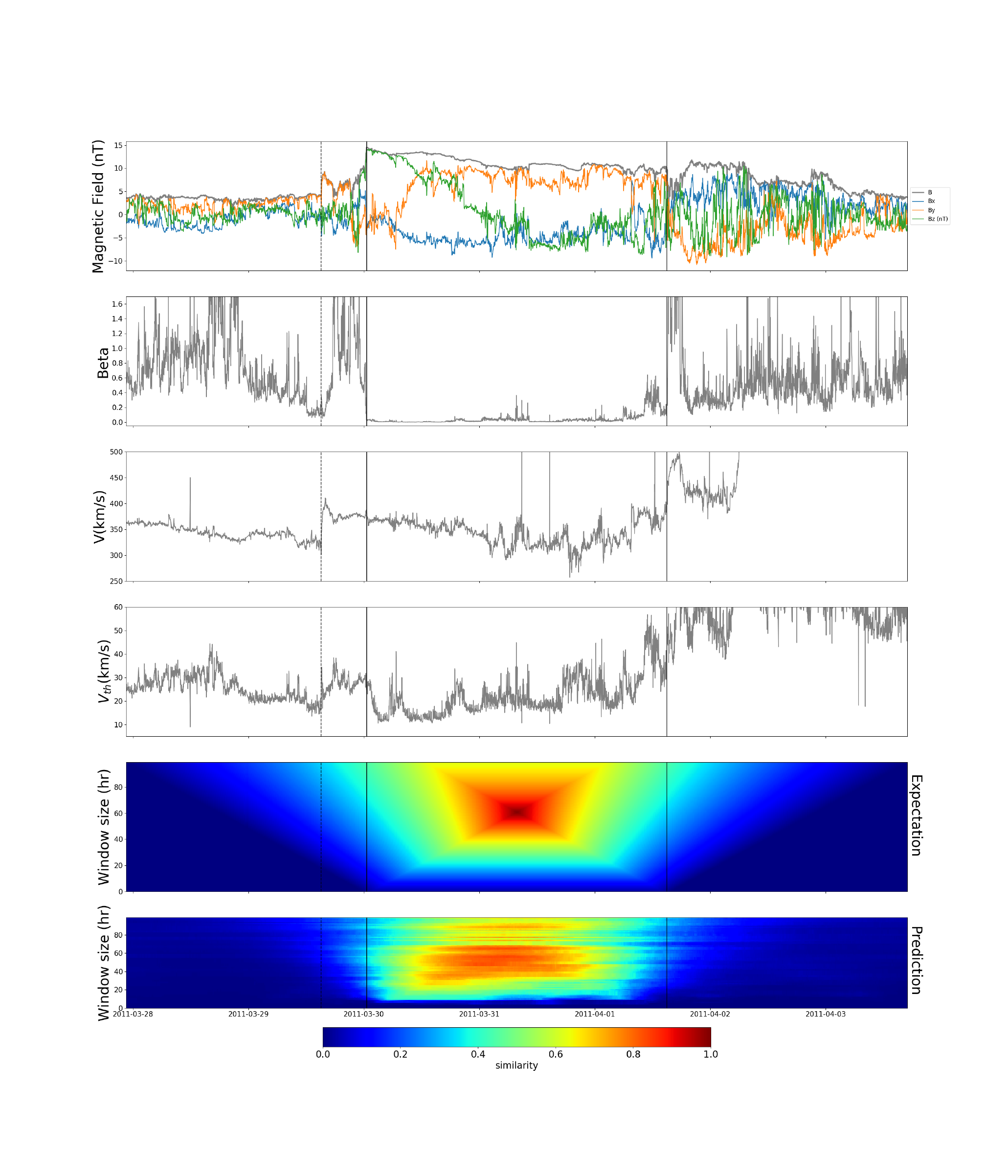}
\caption{Solar wind observation during an ICME from the WIND spacecraft located at the Lagrangian Point L1. The solid vertical lines delimitate the ICME while the dashed vertical line indicate the beginning of the sheath. From the top to the bottom are represented : the magnetic field amplitude and components, the plasma parameter $\beta$, the solar wind velocity, the thermal velocity, the similarity the ICME have with sliding windows of various sizes (from 1 to 100 Hr) and the similarity predicted by our method.}
\label{cloudRepresentation}
\end{figure}

Coronal Mass Ejections (CMEs) are spectacular manifestations of the solar activity which are responsible for the expulsion at large velocties of large quantities of solar plasma and magnetic field. Their interplanetary  counterpart, the so-called Interplanetary Coronal Mass Ejections (ICMEs), interact with the planetary environments. Magnetic clouds (MCs) form a subclass of ICMEs recognized for its strong geoeffectiveness and capable to trigger magnetic storms that severely impact the Earth magnetosphere and ionosphere and even human activities.
After initial studies \citep{Gosling73,Burlaga1981, Burlaga1981Old}, these events have been extensively investigated from in-situ measurements (\citet{KilpuaBible} and references therein). Figure \ref{cloudRepresentation} shows the typical in-situ measurements of a MC from WIND spacecraft which beginning and ending are indicated with the two vertical solid lines. The top panel shows an enhanced magnetic field compared to the surrounding ambient solar wind and a long (Here about 2 days) and smooth rotation of the magnetic field components. It is associated with a low proton temperature (fourth panel) resulting in low values of the parameter $\beta$ defined as the ratio between the kinetic and the magnetic pressure (second panel). The third panel shows an enhanced velocity compared to the preceding solar wind with a declining profile. The MC is featured by abrupt and simultaneous jump in the magnetic field and velocity (indicated by the dashed line in Figure \ref{cloudRepresentation}), and by a sheath, the turbulent region between the shock and the MC. These are the main criteria generally used for the identification of ICMEs \citep{Zurbuchen06, KilpuaBible, Chi16}. However, not all ICMEs feature all standard ICME signatures and there is no signature that would be present in all ICMEs \citep{Gosling73, RC10, KilpuaBible}. For example, about half of the ICMEs drive a fast upstream shock and are preceded by a sheath  (\citet{Chi16} and references therein).

With the launch of the spacecraft WIND and ACE that measure near-Earth solar wind, the number of ICME catalogs associated with these two missions flourished. \citet{Lepping06} referenced 106 Magnetic Clouds (MCs) that are thought to be a special type of ICME between 1995 and 2008. \citet{RC10} listed 373 ICMEs between 1996 and 2015. \citet{Jian06} listed 250 ICMEs between 1995 and 2009. \citet{chinchilla} listed 302 ICMEs between 1997 and 2016. And \citet{Chi16} listed 465 ICMEs from 1995 to 2015.  In average, 80$\%$ of the ICMEs of a given list are present in another list \cite{Chi16}.  The main difference we notice from one list to another stands in the criteria used by the authors to identify the ICMEs such as the presence of a front shock, the presence of a sheath, the fit to a flux rope model \citep{Lepping06} or the importance given by the authors to a specific physical parameter \citep{RC10, Chi16}. Moreover, a given ICME might only fulfill a subset of these criteria or partially fulfill them which complicates their identification and consequently their automated detection.
The establishment of such catalogs allowed the study of ICMEs from a statistical point of view. These studies indicated the enhanced magnetic field and the low proton temperature as being typical characteristic of ICME in-situ signature. These studies also indicated that the yearly occurrence of ICMEs is correlated to the solar cycle \citep{Chi16,Jian06},  and that ICMEs were considered to be long term events with an average duration being equal to 25 hours \citep{Burlaga1981}.  Complete conclusions of such studies can be found in \citet{Chi16}, \citet{chinchilla}, \citet{Mitsakou14}, \citet{KilpuaBible} and references therein.

All the beginning and ending dates of the ICMEs present in these catalogs have been identified by visual inspection. In addition of being quite time-consuming, this task is ambiguous and there are disagreements from an expert to another \citep{Shinde03} because of various reasons, from the criteria considered during the identification to the non-consideration of MCs only, without forgetting psychological factors that can influence the ability an expert has to identify the events. As a consequence, the existing lists are assumed to be incomplete, which partially masks the global vision we could have on the phenomenon. The identification process might even get harder with the growing number of spacecraft providing in-situ measurements of the solar wind and the growing amount of data provided by the totality of these spacecraft. 

The development of an automatic identification method of ICMEs would bring a considerable gain in time and objectivity in the elaboration of our catalogs. \citet{Lepping05} proposed an automatic detection method based on empirical thresholds. These thresholds are inferred from the expert knowledge of ICME properties and involve various physical and temporal parameters such as the duration, the plasma $\beta$, the magnetic field, the bulk velocity or the quality of the fit with a flux rope model. Even though this method was able to recognize a fair quantity of identified events (45 on a total of 76 ICMEs in the period considered), the large number of found false positives (66 for a total of 111 predicted ICMEs) evidenced both the incompleteness of the list as well as the limits of using fixed thresholds for automatic identification. Recently, \citet{Ojeda17} proposed an alternative automatic identification method based on the computation of a Spatio-Temporal Entropy. However, the method was tested on a very low number of ICMEs and its performance on long periods is not known.
Finally, the problem of identifying patterns in in-situ data measurement is at the root of many, if not all observational studies. No matter how efficient previous automatic detection methods were, if some exist, they are based on expert and detailed knowledge of target event properties and thus are very specific to their detection. This methodology thus imposes to re-think the detection pipeline entirely for each kind of event, which constitutes a serious bottleneck that may be comparable or worse than doing the visual identification itself.
\\
One way to overcome these constraints stands in the use of supervised machine-learning algorithms that have the potential to learn the properties of labeled events by themselves.
They represent a promising tool to tackle already large and ever growing bases of reliable data accumulated for decades and their use in space physics is therefore progressing. They have especially been used on solar images for tasks such as the prediction of solar flares (\citet{Colak09} and references therein), the detection of sunspots \citep{Yang18} or even the classification of solar active regions that produced a solar flare with or without a CME \citep{Bobra16}. 
Concerning pattern recognition in time-series, \citet{Pincon96} provided a neural network based method to identify and classify electron and proton whistlers from in-situ data measurements, \citet{Karimabadi07} developed a data mining method called MineTool-TS they used to provide a classification of data intervals that contained Flux Transfer Events (FTE) or not \citep{Karimabadi09} as well as an extension to apply data mining to 3D simulation data \citep{Sipes}. Using a support vector machine on magnetopause crossings measured by 23 different spacecrafts, \citet{Wang13} provided an empirical three dimensional model for the Earth magnetopause. Finally, \citet{Camporeale17} provided an accurate method of solar wind classification into 4 classes using a Gaussian Process. Nevertheless, none of these methods was used to identify the starting and the ending times of a specific kind of event in streaming time series.

In this article, we provide for the first time an automatic detection of the starting and ending times of ICMEs in in-situ streaming data based on deep learning. In section 2, we present the whole pipeline that transforms the in-situ data measured by WIND to a list of predicted ICMEs, in section 3, we will evaluate the performance of our pipeline  based on the number of detected events. Section 4 focuses on the robustness of our pipeline regarding the different dataset features, the number ICMEs used in the training period and the period considered for our training, validation and test phase. Finally, Section 5 will discuss  of the quality of the predicted ICME lists predicted by our model in comparison to experts lists.

\section{Data and pipeline}

\begin{figure}
 \centering
 \includegraphics[scale=0.9]{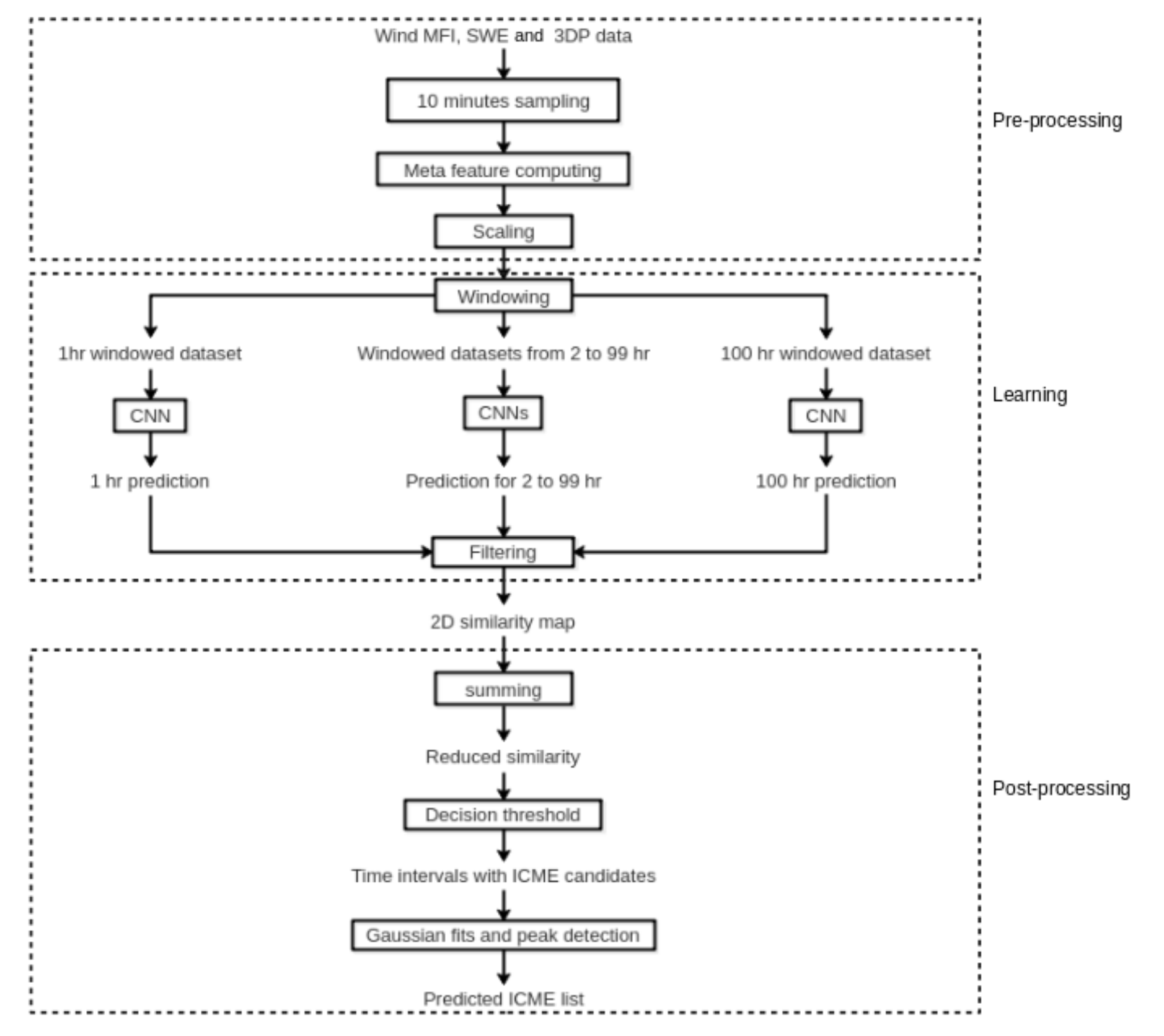}
 \caption{Scheme of our pipeline that converts raw data from WIND spacecraft into a generated ICME catalog, blocks with black contours represents operation lead on the data while blocks without contour represent the state of the dataset at the various steps of the pipeline. The dashed boxes indicate the different major steps of our pipeline. CNN here stands for Convolutional Neural Network, the specific kind of algorithm we will be using in this paper.}
 \label{pipeline}
 \end{figure}

 In this section, we will explain the different steps of the pipeline that  enables automatic ICME detection from WIND in-situ data measurement. A complete representation of our pipeline is shown in Figure \ref{pipeline}. 
 
	\subsection{Data}
The data we used for the elaboration of our method are the data provided by the spacecraft WIND between the 1st of October 1997 and the 1st of January 2016. We used the data provided by the instruments Magnetic Field Investigation (MFI), Solar Wind Experiment (SWE) and 3-D Plasma and Energetic Particles Experiment (3DP), which gave us 30 primary input variables: the bulk  velocity and its components $V,V_{x}, V_{y}, V_{z} $, the thermal velocity $V_{th}$, the magnetic field, its components and their Root Mean Square (RMS) : $B, B_{x}, B_{y}, B_{z}, \sigma_{B_x}, \sigma_{B_y}, \sigma_{B_z}$, the density of protons and $\alpha$ particles obtained from both moment and non-linear analysis : $N_{p}, N_{p,nl}$ and $N_{a,nl}$ as well as 15 canals of proton flux between 0.3 and 10 keV. Due to instrumental constraints, holes are present within the whole dataset, the great majority of these holes have a duration between 2 and 10 minutes. On the other hand, the crossings of ICMEs with their sheath typically have durations of several hours. We therefore resample the data to a 10 minutes resolution, thereby eliminating the greatest majority of the holes while still remaining accurate in the determination of start and end times of labeled events. 

In addition to this 30 input variables, we computed 3 additional features that will also serve as input variables : the plasma parameter $\beta$, defined as the ratio between the thermal and the magnetic pressure, the dynamic pressure $P_{dyn} = N_{p}V^{2}$ and the normalized magnetic fluctuations : $\sigma_{B} = \sqrt{(\sigma_{B_x}^{2}+\sigma_{B_y}^{2}+\sigma_{B_z}^{2}})/B$. In order to evaluate the performance of our method, the dataset has then been split in three parts, the period between 1998 and 2010 constitutes our \textit{training set}, the period between 1997 and 1998 our \textit{validation set} and the period between 2010 and 2016 our \textit{test set}. This repartition has the advantage of considering a whole solar cycle (1997-2008) during the training phase and consequently giving to our algorithm the opportunity to notice the changes in the solar wind and in the frequency of ICMEs during a solar cycle. In section 5 we will split our dataset in two other (train,validation and test) to assess the uncertainty around the quantitative values given in the study.

In the following, this specific configuration of our dataset will be referred as our complete dataset. The data is then scaled and normalized in order to have for each feature an average of 0 and a standard deviation of 1, which prevents a feature preselection due to the difference in orders of magnitude that can exist among them.

\subsection{ICME catalog}

The ICME catalog we used consists in the union of the different WIND ICME lists \citep{RC10, Lepping06, Jian06, Chi16, chinchilla}. During the various tests of our method, additional ICMEs that were not present in any of the existing lists were detected and have been progressively added to our catalog. Following these investigations, 148 new ICMEs have been discovered throughout our dataset period, which represents $22\%$ of our total dataset  for a total of 657 ICMEs distributed as follows: 420 ICMEs in the \textit{training set}, 13 ICMEs in the \textit{validation set} and 232 ICMEs in the \textit{test set}.  
Our catalog can be found online at \url{https://github.com/gautiernguyen/Automatic-detection-of-ICMEs-at-1-AU-a-deep-learning-approach}.
In the following, this catalog will be designed as the Reference List (RL).
We consider that this catalog is still not exhaustive and that events predicted by our pipeline but not being present in the catalog might be in fact actual ICMEs as it will be explained in \ref{subsection:recall}.

\begin{figure}
\centering
\includegraphics[scale=0.5]{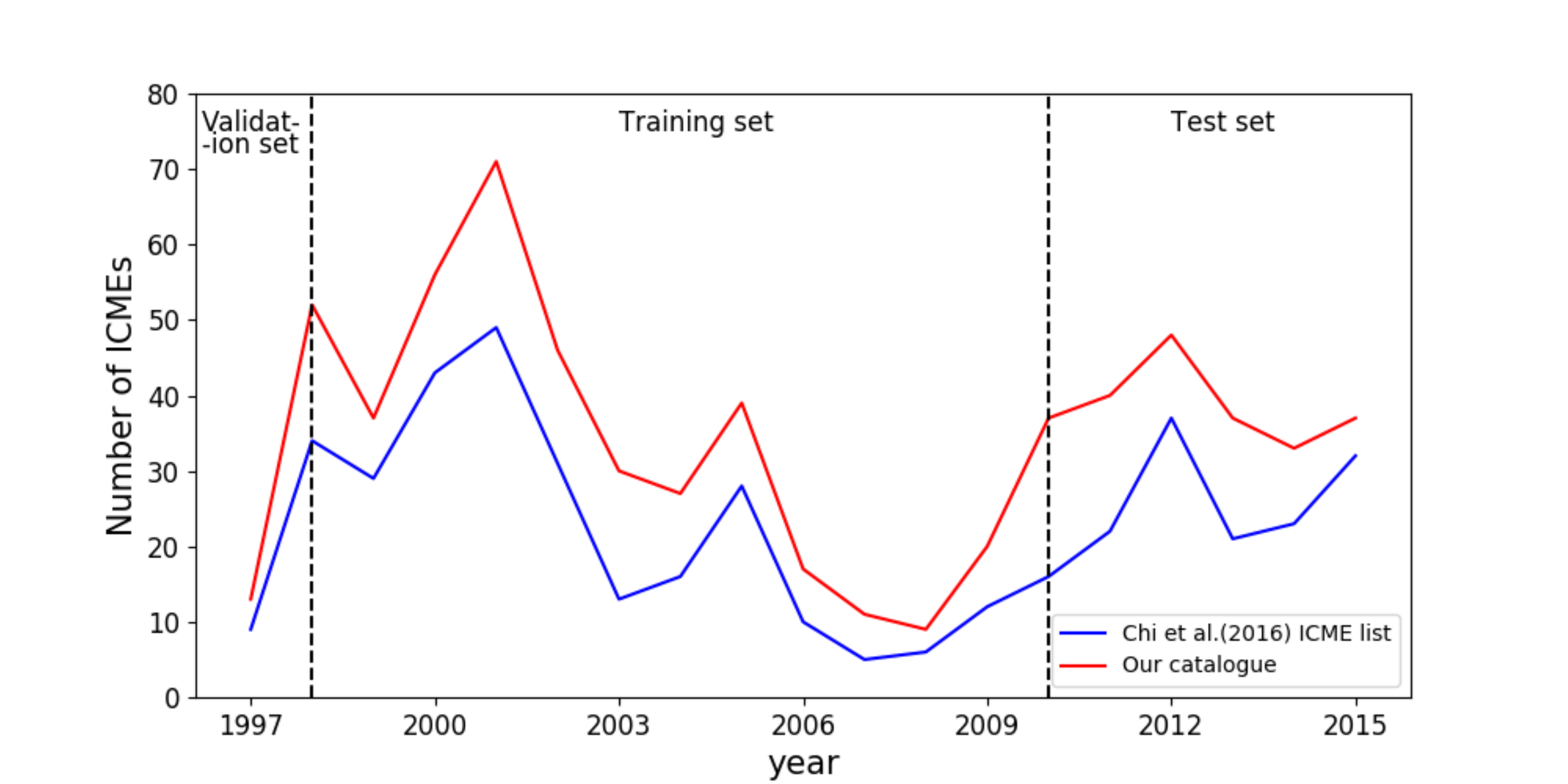}
\caption{Yearly occurrence frequencies of ICMEs of the RL (red) and the list established by \citet{Chi16} (blue), the vertical dashed line indicate the yearly disposition of our training, validation and test set.}
\label{yearRepartLList}
\end{figure}

\begin{figure}
\centering
\includegraphics[scale=0.65]{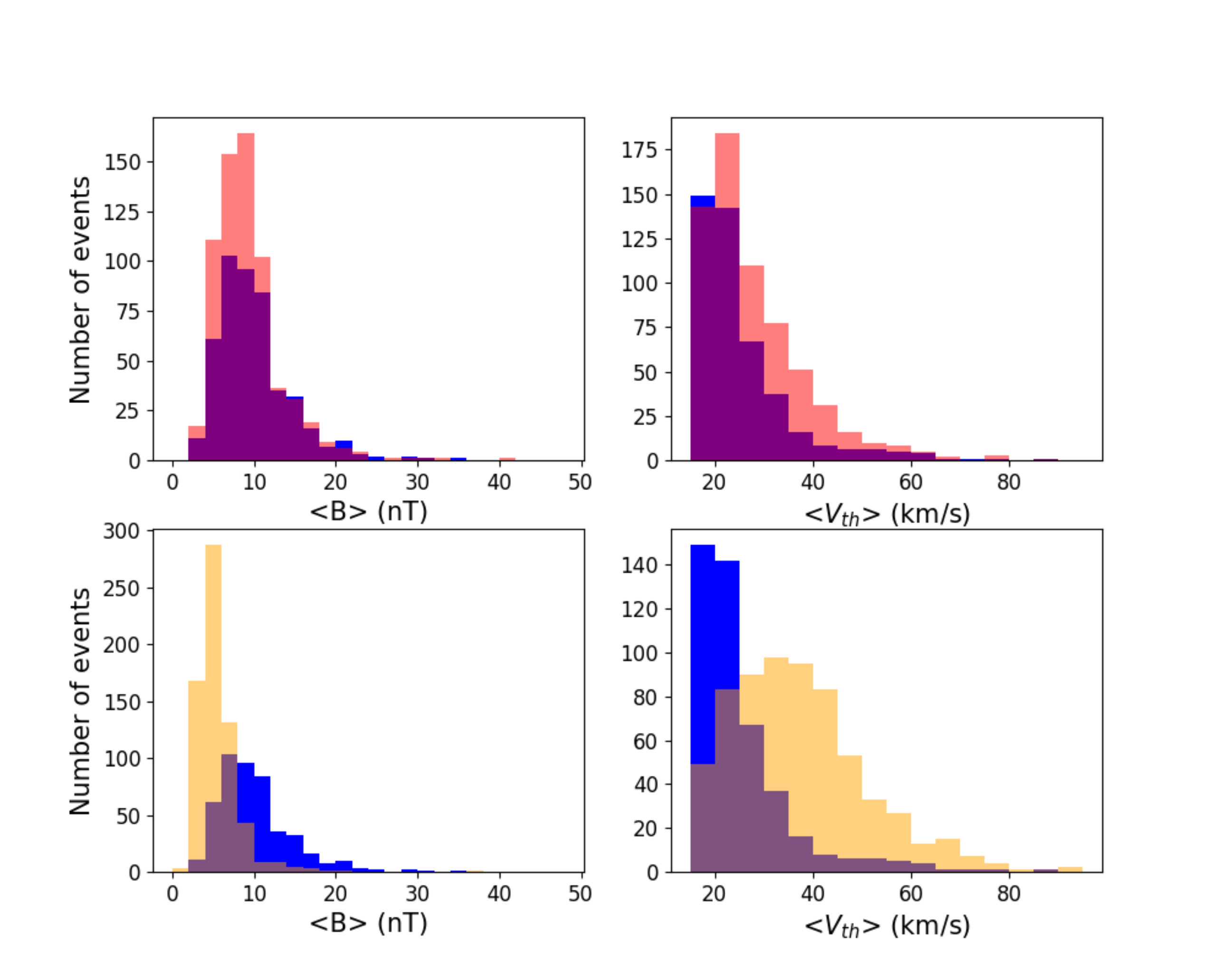}
\includegraphics[scale=0.55]{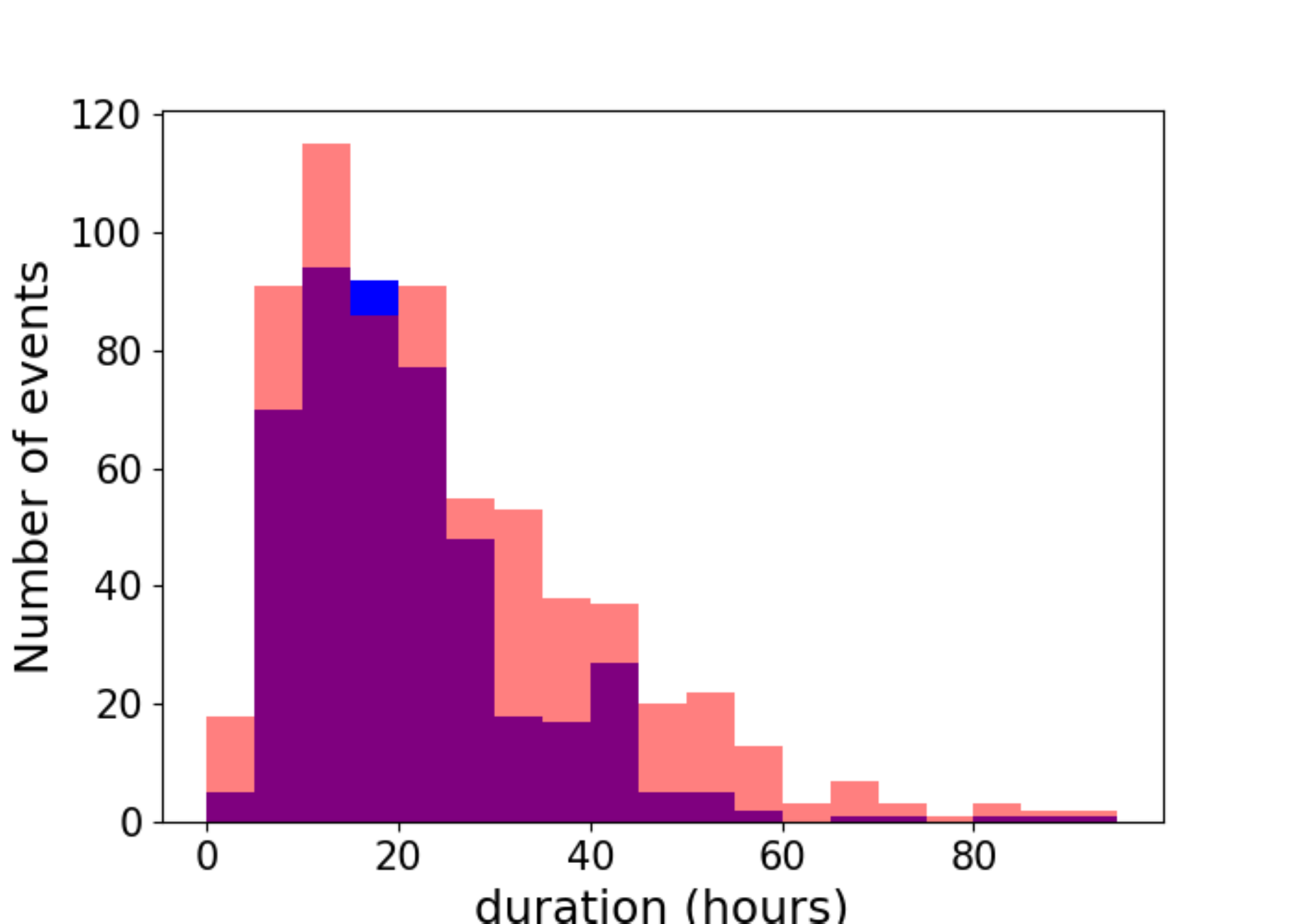}
\caption{First row: distribution of the mean values over the whole ICME interval of the magnetic field and thermal velocity, $\langle B\rangle$ and $\langle V_{th}\rangle$ compared for the list of \citet{Chi16} (blue) and our ICMEs catalog (pink) (overlap in purple). Second row: idem for the list of \citet{Chi16} (blue) and random intervals of solar wind in which there is no ICME (yellow) (overlap in light blue). Bottom Panel : distribution of ICMEs duration for the list of \citet{Chi16} and our ICMEs catalog (same color code as first row).}
\label{histList}
\end{figure}

Statistically speaking, we ensure the consistency of the RL by comparing it to the list established by \citet{Chi16} that has the advantage of being extended over the same time period, as well as being the one containing the most events. Figure \ref{yearRepartLList} compares the yearly occurrence frequencies of the two catalogs. Even if the RL has more ICMEs, the trend observed in the annual variation of the number of ICMEs is conserved and confirms that a whole solar cycle is included in our \textit{training set}. As displayed by the bottom panel of Figure \ref{histList}, the number of events in the RL (in pink) is larger than in the list of \citet{Chi16} (in blue) (the overlap region appears in purple) but with a comparable distribution in duration. The first row of  Figure \ref{histList} also shows consistent distributions of the magnetic field and the thermal velocity between both lists. To ensure this similarity as a proof of consistence, we compare the magnetic field and the thermal velocity of Chi ICME list with random intervals of data in which no ICME was found, the duration of these intervals being distributed according to the duration distribution of our catalog (bottom panel). This comparison in the second row of Figure \ref{histList} shows distributions with larger magnetic fields and reduced thermal velocities in \citet{Chi16} ICME list (in dark blue) than in the list without ICMEs (in yellow) (overlap region in light blue). This is consistent with the expected ICMEs characteristics used by experts for their identification \citep{Zurbuchen06}. On the other hand, this difference compared to the similarity we have for the two ICME lists (first row) ensures that the ICME catalog we used in our identification process is consistent with the previous existing ICMEs catalogs.

\subsection{Windowing and similarity}
 The data is grouped into windows of a hundred different sizes (from 1 to 100hr) that are sliding on both our training and validation sets at a period corresponding to the global dataset sampling : 10 minutes. A window of data represents the values of the 33 input variables within this window that will be treated simultaneously. Our initial dataset is then converted into 100 datasets, each of them corresponding to a size of sliding window. Following this process, there are around 622000 windows of data in the training set, around 311000 in the test set and around 12960 in the validation set. In the following, we will refer to one of these datasets by calling it by its window size.

For each window size, the principle of the detection will stand in estimating a similarity parameter $y_{i}$ for each window of data $X_{i}$, using regression methods for classification purpose. 
Logically, we would expect this parameter to be equal to 0 when no ICME intersects our window while it shall be equal to 1 when a window perfectly matches an ICME.
The similarity $s$ window $X_{i}$ has with a given ICME could then easily be defined by :

\begin{equation}
s(ICME, X_{i}) =\frac{duration(X_{i} \cap ICME)}{duration(X_{i} \cup ICME)}
\end{equation}

Given an ICME list and a window, we then define the expected similarity of the window $X_{i}$ as:

\begin{equation}
s(X_{i}) = \max_{ICME in list} s(ICME, X_{i})
\end{equation}

The aim of our regression would then stand in predicting a similarity $y_{i}$ in order to make it as close to the expected similarity $s({X_{i}})$ as possible.

Stacked together, similarities of many windows make a so-called \textit{2D similarity map}. An example of such a map for a specific ICME is shown on Figure \ref{cloudRepresentation}, fifth panel. The similarity is coded with the color bar, and the ordinate represents the window size from 1 to 100 hours. The maximum is reached in the middle of the event for the window corresponding to the ICME size. One can see that the  similarity  decreases faster in time for small windows than for large windows. Indeed, as they slide along time, small windows cease to see high similarities pretty quickly while large windows remain in range  of ICME-like data - and thus high similarities - for quite longer times.

\begin{figure}
\centering
\includegraphics[scale = 0.6]{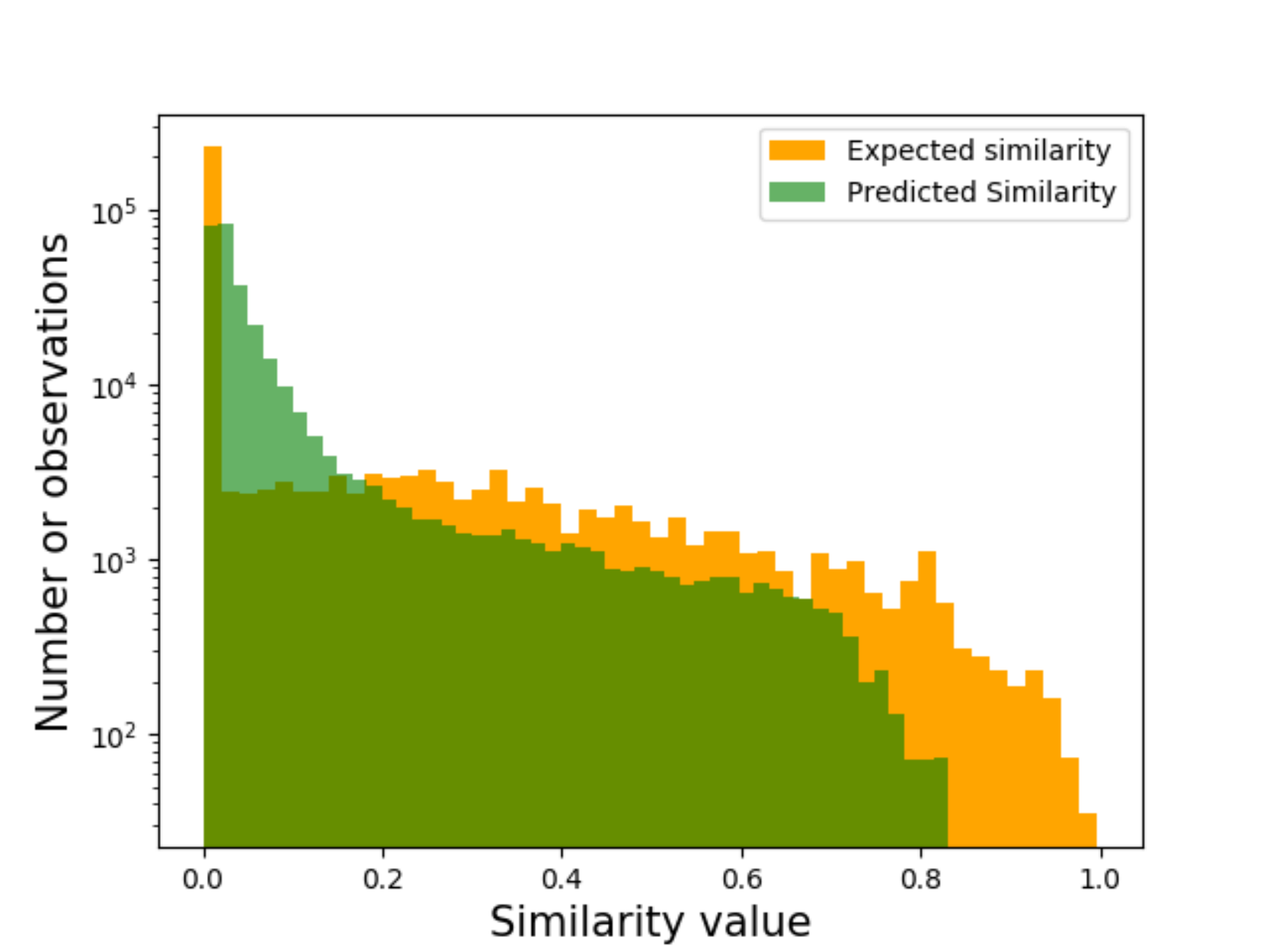}
\caption{Distribution of the similarity values we have on the test set for a window size of 30 hours (orange) and the similarity repartition we have in our prediction for a window size of 30 hours (green).}
\label{simiRepart}
\end{figure}

Figure \ref{simiRepart} represents the distribution in similarity we have on the test set for a window size of 30hr (orange). As the ICMEs are very minority events in the solar wind, the largest part of our computed similarity is going to be equal to 0.

	\subsection{Algorithm}

For each window size, we use a Convolutional Neural Network (CNN) to learn and then predict the similarity of a given window associated with the underlying data.
CNNs represent a class of neural networks mostly used for computer vision or image classification. Like every other kind of deep neural network,  they consist of an arrangement of interconnected nodes, or neurons, forming a certain number of layers that process the information in a way reminding the human brain. In the case of a CNN, the neurons consist of learnable filters which objective is to detect features or patterns present in our input data. Such filters are convolved (slided) along the input data and indicate the presence or not of the requested pattern through an activation function. A complete description of the principles of CNNs can be found in \citet{Geron}. The architecture we used for our CNN is similar to the one proposed by \citet{Yang2015} that had been used to recognize human activities from multi channel time series. The exact architecture of the CNNs we used can be found in appendix A. 
\\
 The training (resp. prediction) phase is done by sliding windows over the data to learn (resp. predict) the similarity from data. Different window sizes are used, each corresponding to the training or prediction of a specific CNN. The \textit{training set} defined previously is used to determine the characteristics of each filter present in the convolutional layers in order to minimize a cost function $J$ of our training set that quantifies the average error made by the algorithm.Starting with no initial knowledge about ICMEs nor any kind of event, likely to appear in in-situ data measurement, the initial characteristics of each filters were initially set with random values. Due to the imbalance of our problem it will be difficult for our prediction to reach high similarity values. In order to be less sensitive to the errors we would make on these high similarity values, we define $J$ as:

\begin{equation}
J(\textbf{X}) = \frac{1}{n} \sum\limits_{i=1}^n \log [\cosh(y_{i}-s(X_{i}))]
\end{equation}

where $X_{i}$ denotes a window of data that contains the values of each features during the time the window is defined, $\textbf{X}=\lbrace X_{i}, i \in [\![ 1;n ]\!] \rbrace$ represents the windows in their globality and $y_{i}$ is the estimated similarity for the window $X_{i}$. To ensure the capacity the algorithm has to generalize on unknown data, we estimate the similarity on the \textit{validation set} and train our algorithm again using the filters characteristics we previously obtained. This process is repeated for a 100 iterations, or epochs, unless the computation of $J$ over our \textit{validation set} reaches its minimum before. In addition to avoiding the overfit, this process has the advantage of bringing the minimization closer to the minimum of $J$ and, thus, to provide a finer estimation of the similarity.

As shown in Figure \ref{simiRepart}, the prior we give to our algorithm is unbalanced, having too many values being equal to 0 and not enough high values. To cope with this imbalance, each window of a given size is weighted. Windows having a similarity of 0 will keep a weight of 1 while the windows of other values will be granted a weight corresponding to the inverse of their occurrence frequency in the similarity distribution. The distribution of the predicted similarities for a window size of 30 hours is also shown in figure \ref{simiRepart} (in green). As less importance was given to the window for which the similarity was 0, the similarity of a lot of windows was estimated to be a little bit higher than 0, this explains the high number of windows having a similarity between 0 and 0.2 in the Figure \ref{simiRepart}. Moreover, even if a particular attention has been set on the windows having high values of similarity, their number is still very low and consequently high similarity windows are not seen enough by the algorithm during the training phase. This explains why our predictions does not manage to estimate similarity up to its maximal value. Nevertheless, the  weighting of windows allowed the estimation of high values of similarity and consequently allowed a finer detection of ICMEs.

We then used our algorithms on the \textit{test set} (corresponding to the 2010-2016) period to evaluate the capacity our algorithm to detect ICMEs on unknown data. For both our training and prediction phase, we used two NVIDIA GeForce GTX 1080 TI $\copyright$ in parallel. The training phase for our 100 CNNs usually lasted 35 hours while the prediction for our 100 windows over 6 years usually lasted 50 minutes, which allows us to generate a list of ICMEs in a very short amount of time compared to human labelling.

    \subsection{Post-processing}
\begin{figure}
\centering
\includegraphics[scale=0.6]{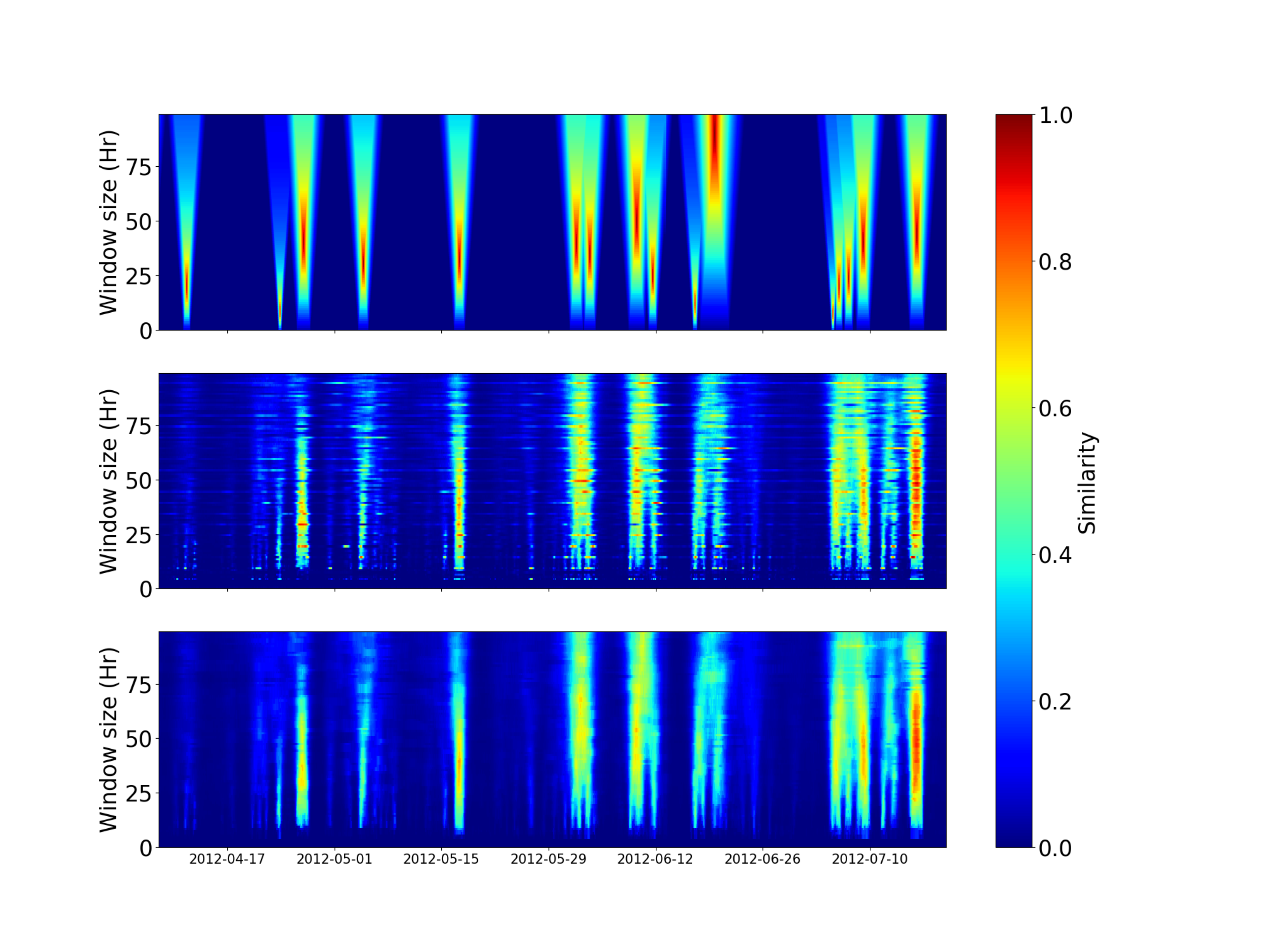}
\caption{Three different representations of the similarity parameter during the period between 8 April 2012 and 20 July 2012. \textit{Top} : Theoretical similarity computed from our prior ICME list, colored regions correspond to an actual ICME of our catalog. \textit{Middle} : raw predictions obtained from each of our CNN. \textit{Bottom}: Actual prediction of the pipeline after applying a median-filter to the raw predictions.}
\label{predictedSimi}
\end{figure}

The first panel of Figure \ref{predictedSimi} represents the expected similarity coded with the color bar computed for each of 100 CNNs (windows)  during the period between the 8th of April 2012 and the 20th of July 2012. The raw predictions of the similarities (second panel) for each of these 100 CNNs are remarkably consistent with the expected similarities. However a small noise is present in this raw prediction, which is thus smoothed using a median filter, and gives us the third panel of Figure \ref{predictedSimi}. It is worth noting already from this figure that the high predicted similarity values are very well localized in time on intervals very close to those of cataloged ICMEs, and there is a quite high contrast with the ambient solar wind background. Additionally, the predicted similarity seems to be high for intermediate window sizes and lower for small and large windows, indicating the algorithm has well learned the typical length of ICMEs from the training set. In addition, it is remarkable that ICMEs that are quite close from each other are not well separated by large windows, but often are separated by smaller ones. This is quite reasonable since, to the CNN, large windows features roughly look like a single ICME, whereas small windows actually have a chance to see data intervals that do not. The CNN here faces the same dilemma an observer would: "does an ICME cover the whole 100 hours, or are there two ICMEs, one closely following the other?". In these cases of true observational ambiguity, our 2D similarity map does not choose for the observer but rather provides a multi-scale  suggestion.
The smoothed predicted similarity is also shown on the bottom panel of Figure \ref{cloudRepresentation} for an interval zoomed over a single ICME. Like the expected similarity, one can see a faster decrease of the similarity for small windows than for the large one. It is worth noting that plotting the predicted similarity together with data is of great interest to very quickly  and unambiguously identify visually ICMEs or periods of ICMEs, thereby shortening the usually long phases of data selection for observers. For some events, that will be qualified later as false negatives, the CNN has failed to predict a high similarity. This is the case of the first ICME of Figure \ref{predictedSimi}. Interestingly though, 2D maps still reveal a weak but coherent signature over the 100 - independent - CNN predictions, and that is visually detached from the ambient solar wind zero similarity. Similar weak and coherent similarity signatures are sometimes seen where no ICME is labeled in the catalog. For observers, 2D maps are only truly useful if they highlight a large percentage of the ICMEs, and do not wrongly highlight too many intervals with no ICME in it, so-called \emph{false positives}. An algorithm having these two properties is said to have a good recall and a good precision, respectively. These properties will be estimated quantitatively in the next section, to evaluate the performance of our detector.\\

\begin{figure}
\centering
\includegraphics[scale = 0.55]{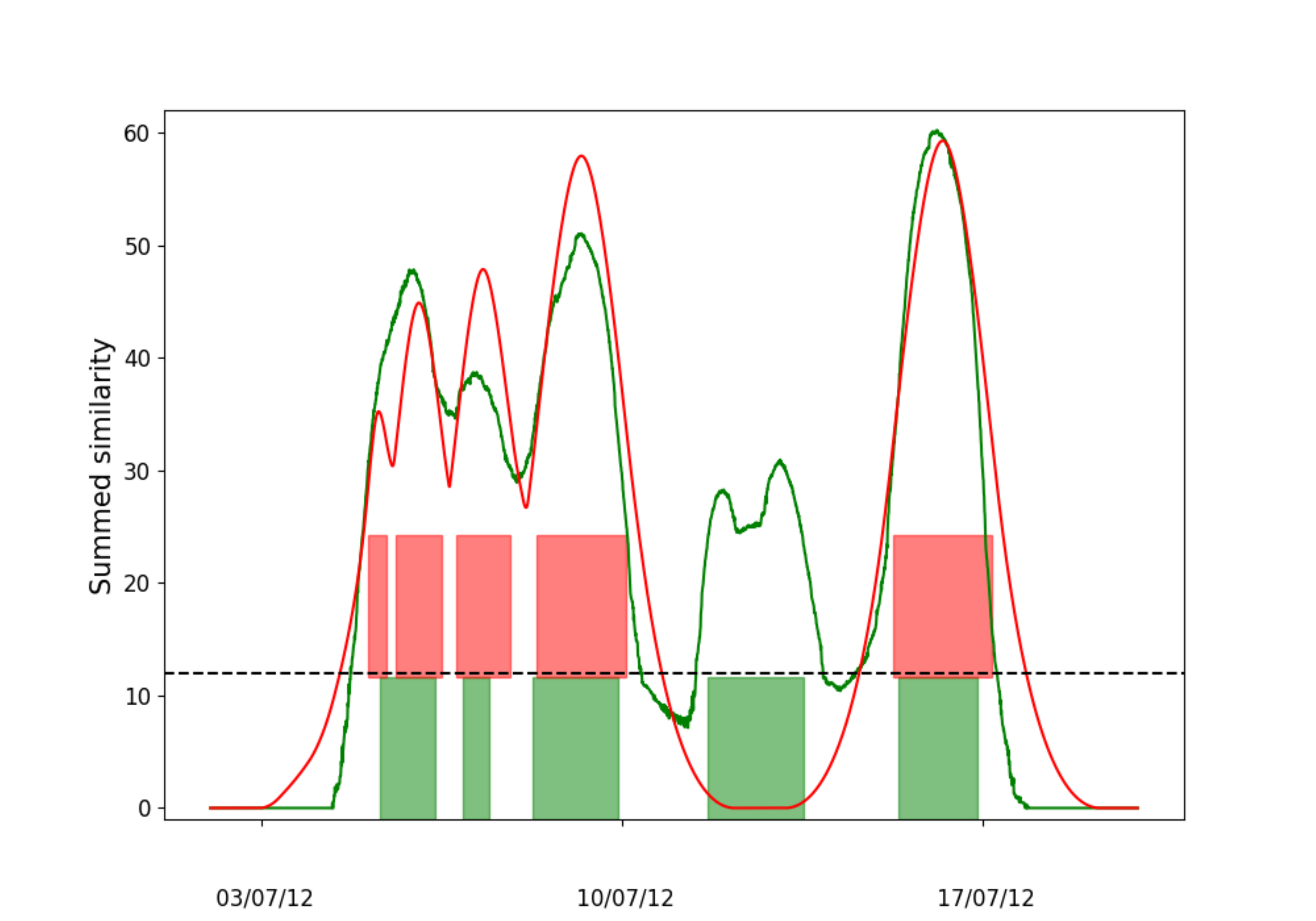}
\caption{Expected (red) and predicted (green) reduced similarities for the period between the 3rd July 2012 and the 17th of July 2012. The red regions correspond to expected ICMEs from our catalog. Green regions correspond to predicted ICMEs after applying gaussian fitting and peak detection. The dashed line indicates the decision threshold (here equal to 12) we choose to make our prediction. The performances associated to this threshold correspond to the black dot of Figure \ref{precisionRecall}.}
\label{GaussianFits}
\end{figure}

\subsection{Automatization}
\label{subsection:automatization}
The evaluation of the detection performance requires an automatization process that converts our predicted similarity into a list of predicted ICMEs. An important by-product of this process is the production of a reproducible and objective ICMEs catalog that can thus easily be updated incrementally with time without human intervention.
Because we only want to generate a list of start and end times, and because of noise in the window size dimension persisting after the application of the median filter, we reduce the predicted similarity to its time dependent integral along the window size axis, which defines the so-called \textit{reduced similarity}. We then regularize it by using a multiple-gaussian fit. An example of the fitted reduced similarity is shown as the green curve on Figure \ref{GaussianFits} for the period running from July 3rd 2012 to July 7th 2012. A primary criterion is then applied to determine intervals within which ICME candidates may be searched. These intervals are defined as those for which the similarity exceeds a so-called \emph{decision threshold}, shown as the dashed horizontal line in Figure \ref{GaussianFits}. Finally, a peak detection algorithm is applied to each of these intervals, from which the peak times and the half-height times will define our center, start and end times for predicted ICMEs. To deal with the possible ambiguity that can exist for two close-by events, neighboring predicted ICMEs by less than two hours will be merged and considered as a single predicted event This is the case for the fourth predicted ICME of Figure \ref{GaussianFits}. Finally, a predicted ICME having a duration of less than 2 hours is automatically considered as an inconsistent prediction since the CNN has never learned such a short duration event from our list, and is removed from the predicted list. Predicted ICME intervals are represented on Figure \ref{GaussianFits} as green rectangles. The post-processing part of the pipeline is also summarized on Figure \ref{pipeline}.

\section{Results}
\label{section:performance}

The performance of our method simply consists in comparing the predicted ICME list, obtained as explained above, with the ICMEs of the RL that are in the test period. It is important to remind that the RL, as any catalog, is not exhaustive and there are still ICME-like intervals never labeled in WIND data. Furthermore, time series represent a one dimensional slice into a non-stationary three-dimensional structure,  therefore start and end times isolating events are based on an interpretation of the data. Labels do not represent an absolute truth, they vary from one expert to the other and one cannot expect any algorithm to outperform human in this subjective task. As a consequence, performance metrics are not perfect, their estimate cannot be expected to reach 100\%.

\subsection{\sout{Performance evaluation} Precision and Recall}
 Test ICME intervals are shown on Figure \ref{GaussianFits} as red rectangles. An existing (red) ICME is then considered as detected if more than $50 \%$ of its duration is overlapped by a predicted (green) ICME.
Due to the possible ambiguity that can exist in the transition from an ICME to a neighboring one, two ICMEs of our catalog are allowed to be detected by the same predicted ICME, which is the case for the first predicted ICME of Figure \ref{GaussianFits}.

From then on, we can separate the errors made by our method into two categories:

\begin{itemize}
\item A false negative (FN) is an ICME from the RL that is not detected by our method
\item A false positive (FP) is an ICME predicted by our method that is not in the RL
\end{itemize}

These values are obtained for a given decision threshold.
Logically, the low values of this threshold allow weak similarity peaks to be seen as predicted ICMEs, thereby increasing false positives but decreasing false negatives. Inversely, high decision thresholds result in less false positives but more false negatives. The so-called \emph{precision} $P$ of the algorithm, is defined as one minus the ratio of FPs and the total number of predicted ICMEs $N_{pred}$.

\begin{equation}
P = 1 - \frac{N_{FPs}}{N_{pred}}
\end{equation}

The so-called \emph{recall} $R$ of the algorithm, is defined as one minus the ratio between the FNs and the total number of ICMEs in our test set  $N_{expected}$ (here being equal to 232)

\begin{equation}
R = 1- \frac{N_{FNs}}{N_{expected}}
\end{equation}

\begin{figure}
\centering
\includegraphics[scale = 0.5]{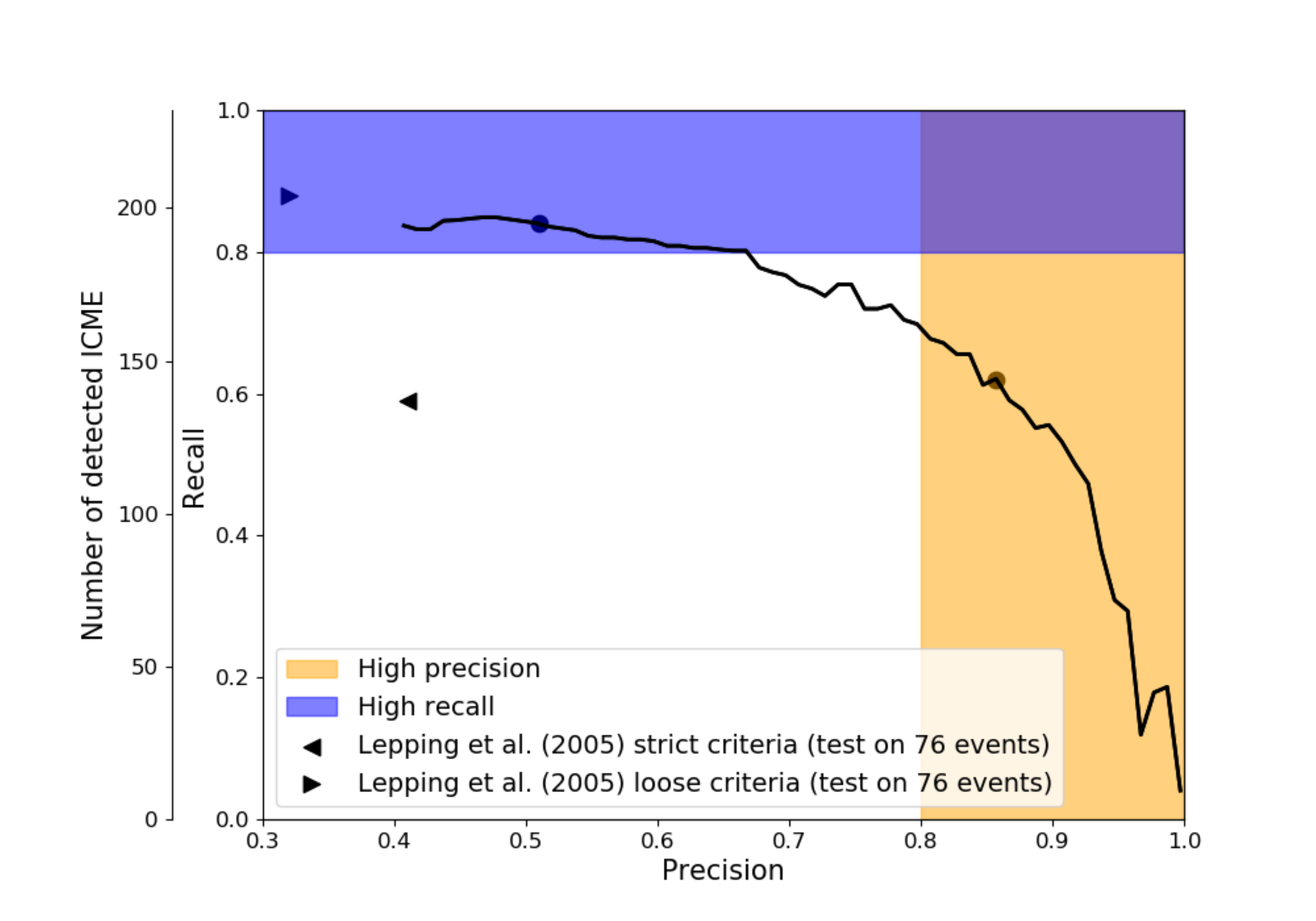}
\caption{ Precision-recall curve (black line) of our method. The region in blue indicates zones of high recall and fair precision where flux ropes and small ICME-like events can be detected. The region in yellow maps the high-precision zone for which the greatest part of the predicted ICMEs are in our catalog but do not represent it fully. The three markers indicate the performances of the previous attempts of ICME automatic identification : \citet{Lepping05} strict (leftward triangle) and loose (rightward triangle) criteria. The black dot in the high precision region is associated with the decision threshold presented in Figure \ref{GaussianFits} and with the discussion of subsection \ref{subsection:precision}. The black dot in the high recall region represents the working point used in the discussion of subsection \ref{subsection:recall}.}
\label{precisionRecall}
\end{figure}

Computing the recall and precision for a continuously varying decision threshold gives us the evolution of the precision as function of the recall, as represented in the precision-recall curve in Figure \ref{precisionRecall}. As expected, an increasing precision goes with a decreasing recall. The irregularities that are found on the curve, especially for high precision can be explained by the total number of predicted ICMEs that changes when we change our decision threshold. High recall and high precision regions are shown on Figure \ref{precisionRecall} as colored rectangles, the performances of our method in these regions will be detailed later-on. A perfect algorithm works at recall and precision 1. In practice it is rarely the case and observers will need to adjust the decision threshold so to maximize the recall, at the price of a smaller precision, or vice versa, depending on the objective. For physical studies, it seems to us that one may prefer high precision at the price of less events being detected provided the method still detects a fair number of events. Figure \ref{precisionRecall} shows recall and precision obtained in a previous work proposing automatic ICME detection  \citep{Lepping05}. Because this method consists in fixing several arbitrary thresholds, it cannot easily produce precision recall curves, and only few points are to be compared with. Unlike it, our method comes with a single handle for the decision threshold, that moreover does not require prior knowledge of the physical nature of the events to be detected. Let us note here that at equivalent values of recall (resp. precision), our method leads to higher precision (resp. recall) and even reaches values of precision and recall that had not been reached by automatic identification methods yet.

To ensure the capacity our algorithm has to generalize on unknown data we also made tester our whole pipeline on the training set and compared the performances of both predictions. To do so, we compute for both predictions the area under the curve of the precision-recall associated with the predictions.  To cope with the changing minimal precision of each dataset, the computed integral  has  been  normalized  by  the  area  of  the  zone  delimited  by  the  minimal  precision  reached  during  the  post process.  For an ideal identification method, this value of the so-called \textit{average-precision} would be 1.  Since our problem is limited by the ambiguity we have on the starting and ending times of the ICMEs signature combined with the non-exhaustivity of the RL, the average precision on our training set is 0.743 while it is 0.697 on the test set. As the training set has been used to set the characteristics of the filters of the CNNs, it is not surprising to see a higher score on this prediction.   The two scores are in the same order of value, which proves the capacity our algorithm has to generalize the knowledge about ICMEs it learned to unknown data.

\subsection{High recall region}
\label{subsection:recall}
\begin{figure}
\centering
\includegraphics[scale=0.55]{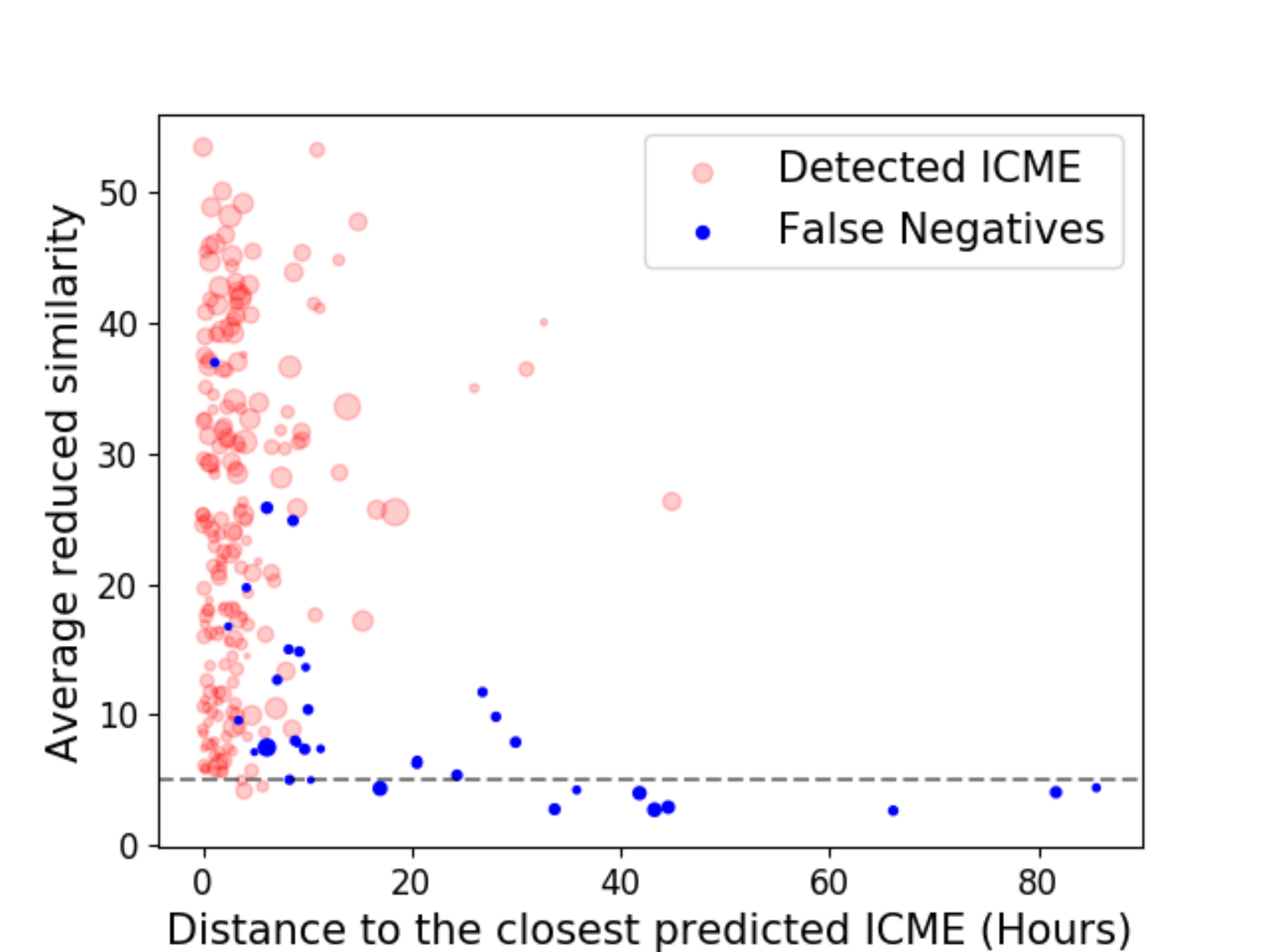}
\caption{Average reduced similarity for each ICME of our test period as a function of the temporal shift to the closest predicted ICME. Red corresponds to detected ICMEs and blue corresponds to the FNs. The dashed line indicate the decision threshold we chose to make the prediction. The size of the circles corresponds to the duration of the event.}
\label{scatterFNhighRecall}
\end{figure}

The high recall region is represented by the blue rectangle in the Figure \ref{precisionRecall}. In this region, our method detects the greatest part of the ICMEs present in the test set while generating a fair number of false positives. To quantify the performances of our method, we selected the working point shown in Figure \ref{precisionRecall}. At this point, 197 of the 232 ICMEs present in our test period are detected, which means a total recall of 84$\%$ and 160 of the 330 predicted ICMEs are considered as false positives, meaning a total precision of 51$\%$. The difference we notice between the number of detected ICMEs and the number of predicted ICMEs that are not false positives (170) comes from the fact that we allow neighboring ICMEs to be detected by the same predicted event. For this decision threshold, 136 of the 150 ICMEs present in the list of \citet{Chi16} were detected, which represents a recall of 91 $\%$ on this list. Likewise, we detected 100 of the 111 ICMEs present in the list of \citet{chinchilla} with a recall being equal to 90$\%$.
\\
As our recall is close to its maximal value, the FNs we have correspond to ICMEs that will not be detected by this model whatever our decision threshold is. It is then in our interest to characterize these events. For the decision threshold we chose, we obtained 35 FNs, 10 were exclusive to the list of \citet{Chi16}, 7 came from the list of \citet{chinchilla}, 3 were common to both lists and the 15 remaining came from the ICMEs we added after the different tests of our pipeline. In the  Figure \ref{scatterFNhighRecall}, we represented the average reduced similarity of each ICME as a function of the time shift with the closest predicted ICME (that can also be a FP). The red points represents detected ICMEs while the blue one represent FNs. The size of each point being proportional to the associated ICME duration. The dashed line represents the decision threshold we chose to make our prediction. For a detected ICME, the closest predicted ICME is expected to be as close to the ICME as possible. This is why the time shift is usually low for detected ICMEs. The reason for which we can have more than 15 hours in the shift stands in the possibility we have to merge the predicted ICMEs. Additionally, it is not surprising to notice that the greatest majority of the detected ICMEs have an average reduced similarity above the decision threshold. Concerning the FNs, we can split them in two categories. On the one hand, 11 of them have an average reduced similarity below the decision threshold and can in fact be considered too weak in both their duration and pattern to be detected. On the other hand, all of the FNs with an average reduced similarity above the threshold are distant from a predicted ICME by less than 30 hours. Because of this proximity, the pipeline might not be able to distinguish the transition from an ICME to another and the prediction of the FN is then \textit{absorbed} by its closest neighbor. Additionally, all of the FNs appear to have a short duration. Consequently, our FNs are short ICMEs that are either too weak to be detected, or close to another predicted ICME that entails their detection.

\begin{figure}	
\centering
\includegraphics[scale=0.6]{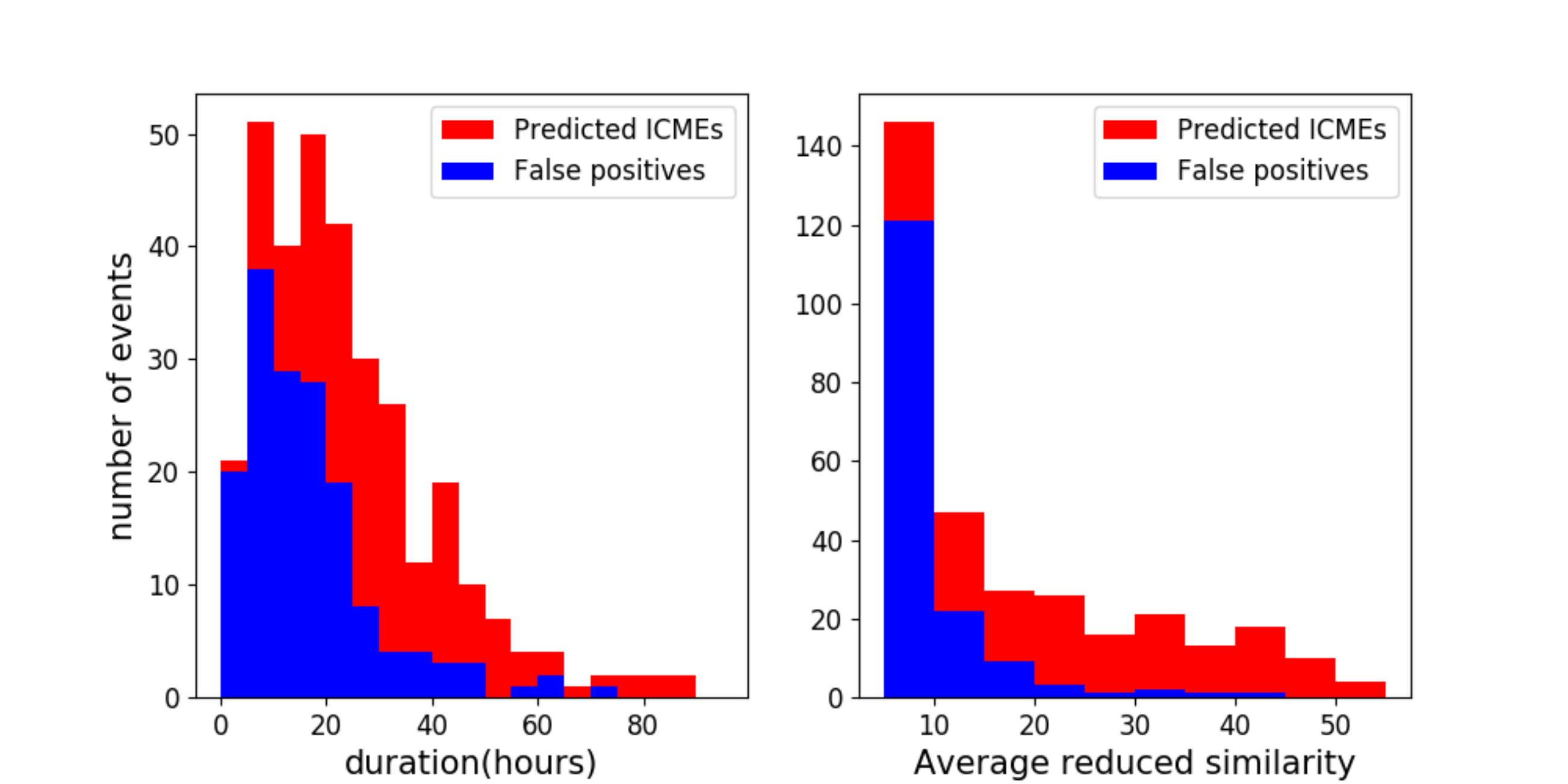}
\caption{Distribution of the duration (left) and the mean value of the integral (right) for the predicted ICMEs (red) and the false positives (blue).}
\label{histFPhighRecall}
\end{figure}

Similarly, a characterization of our FPs is necessary to understand the origin of the errors made by the model. Additionally, we assume the RL to be not exhaustive, investigating these FPs would then be the opportunity to potentially discover ICMEs that had not been discovered yet. Figure \ref{histFPhighRecall} represents the distribution of the ICMEs predicted by our pipeline according to their duration and the mean value of the reduced similarity during the event. As the FPs are a subset of these predicted ICMEs, the red bins then represent the predicted ICME that do contain one or several ICME of our catalog. The possibility a predicted ICME has to cover several expected ICMEs is also at the origin of the difference we notice with the duration distribution shown in Figure \ref{histList}. The left panel shows the great majority of the predicted ICMEs having a duration below 20 hours are FPs. Similarly, the right panel shows that most of the ICMEs predicted with a low mean value of the reduced similarity are actually FPs. This confirms that augmenting the decision threshold is a good way to drastically reduce the number of FPs. Most of our FPs appear to have a short duration and a low mean value of the reduced similarity. An efficient way of ignoring them would then stand in the establishment of criteria according to these two parameters. However, as our catalog is assumed to be not exhaustive yet, some of these FPs might in fact be regions in the dataset where there could be one or several ICMEs that have not been discovered yet.

\begin{figure}
\centering
\includegraphics[scale=0.18]{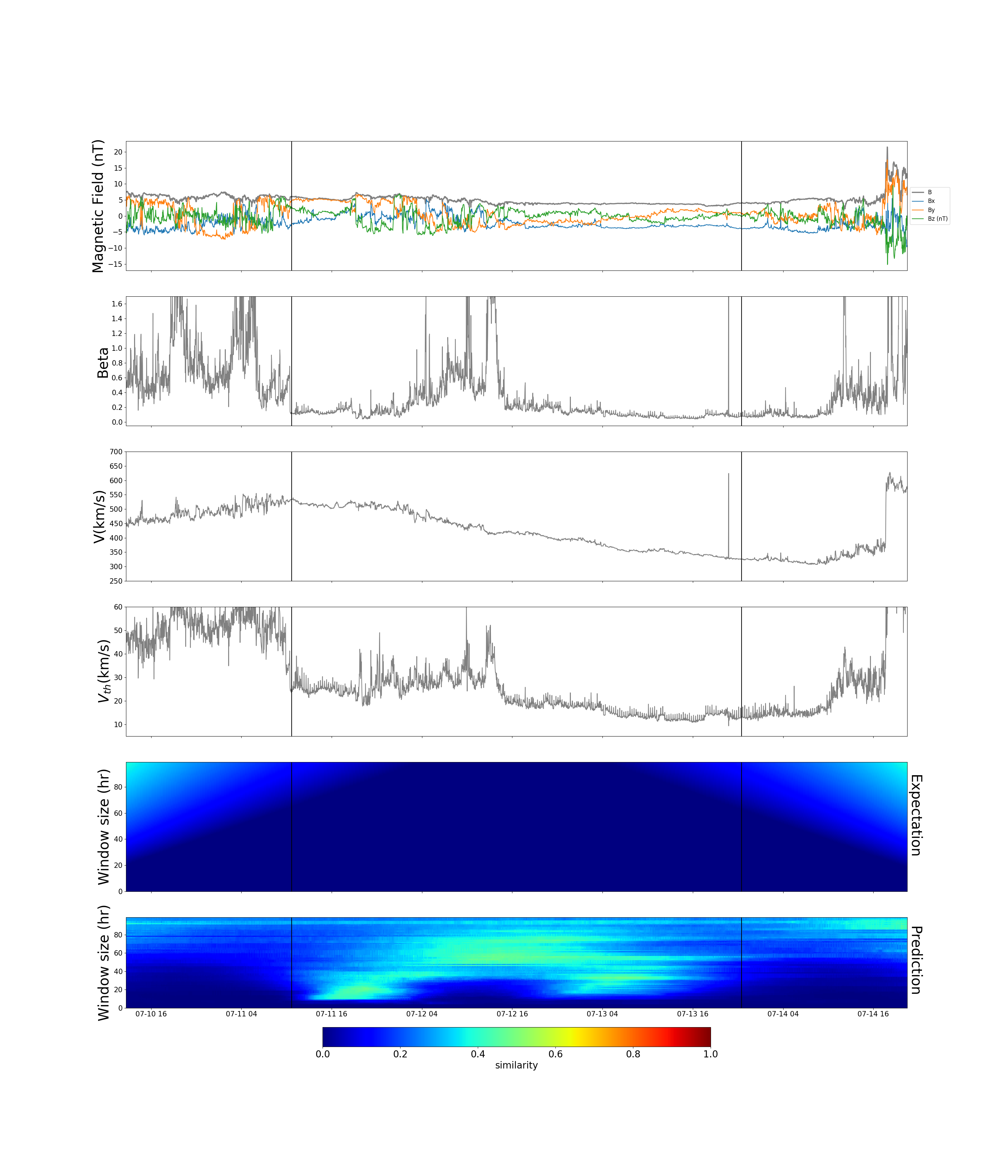}
\caption{Solar wind observation of the FP predicted by our pipeline in the figure \ref{GaussianFits} that contains two ICME like events. The solid vertical lines delimitate the boundaries of the predicted event. From the top to the bottom are represented : the magnetic field amplitude and components, the plasma parameter $\beta$, the solar wind velocity, the thermal velocity, the similarity the ICME have with sliding windows of various sizes (from 1 to 100 Hr) and the similarity predicted by our method.}
\label{icmeLike}
\end{figure}

\begin{figure}
\centering
\includegraphics[scale=0.6]{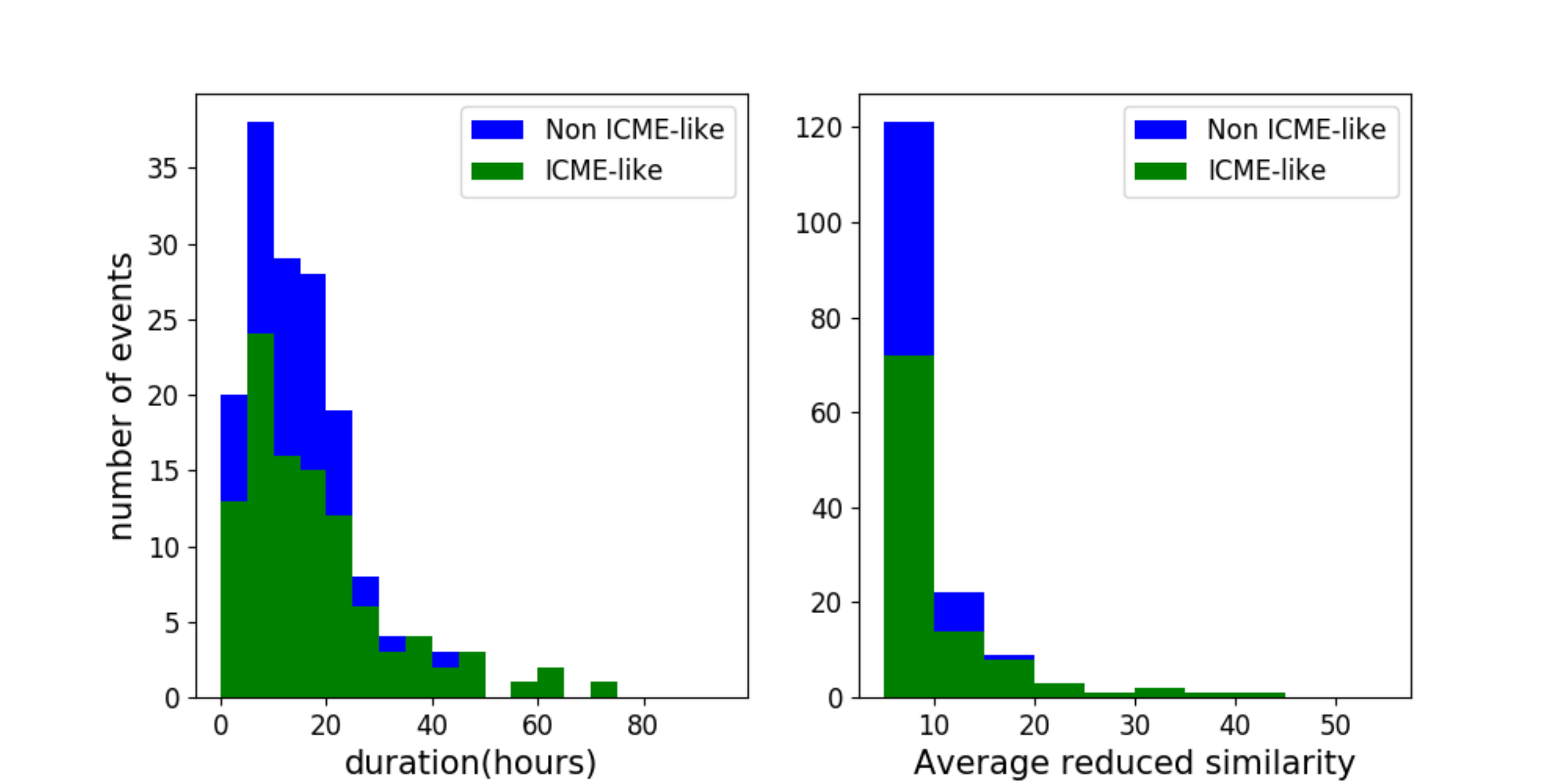}
\caption{Distribution of the duration (left) and the mean value of the integral (right) for the FP depending if they are ICME-like (green) or not (blue).}
\label{histICMELikehighRecall}
\end{figure}

For these reasons, we inspected visually the 160 supposedly FPs and made a distinction between the \textit{ICME-like} FPs, that did contained one or several time intervals that were susceptible of being actual ICMEs, and the \textit{non ICME-like} FPs. A FP was considered as \textit{ICME-like} if fitted at least three of the criteria used by \citet{Chi16} to identify ICMEs manually. 102 of our 160 FPs were considered to be ICME-like while 58 others were considered to be non ICME-like. Figure \ref{icmeLike} represents the in-situ observation of the FPs predicted by the models in the Figure \ref{GaussianFits} between the 11th of July 2012 and the 14th of July 2012 with the same panels than the one exposed in Figure \ref{cloudRepresentation}. Between the two vertical solid lines that indicate the boundaries of the FPs, one can see two ICME like regions. This appears very clearly on the 2D maps for predicted similarity (bottom panel in Figure \ref{icmeLike}) with a single spot for high window sizes that splits into two for the low window sizes. In this case, the FPs appear to have a large duration and a high mean value of the reduced similarity. One could then wonder if these two parameters could also serve as criteria to discriminate the ICME-like from the non ICME-like among the FPs. Figure \ref{histICMELikehighRecall} 
represents the distribution of the FPs predicted by our pipeline according to their duration and the mean value of the reduced similarity during the event. The blue bins represent the non ICME-like FPs while the green bins represent the ICME-like. Even if there is no real temporal discrimination visible on the left panel, FPs having a duration higher than 20 hr are likely to contain one or several ICME-like regions while the non-ICME like FPs usually have a short duration below 20 hr. Looking at the right panel, the reduced similarity then appears as a useful parameter to identify ICME-like FPs. Indeed, the great majority of the non ICME-like appear to have the lowest mean reduced similarity values while all but one FPs having the highest mean reduced similarity values are in fact ICME-like.

Setting our decision threshold in order to be in the high recall zone would then be useful to detect additional ICME in order to complete our current catalog. During the numerous trials of the pipeline, we then regularly checked the FPs predicted by our models in order to complete our catalog with potential new ICMEs. In order to find new ICMEs in the whole 1997-2015 period, predictions were also made on the 1997-2003 and 2004-2009 period by using the remaining period of the dataset for the training and the validation of our pipeline. This investigation led us to the ICME catalog that was presented previously.

\subsection{High precision region}
\label{subsection:precision}
\begin{figure}
\centering
\includegraphics[scale=0.4]{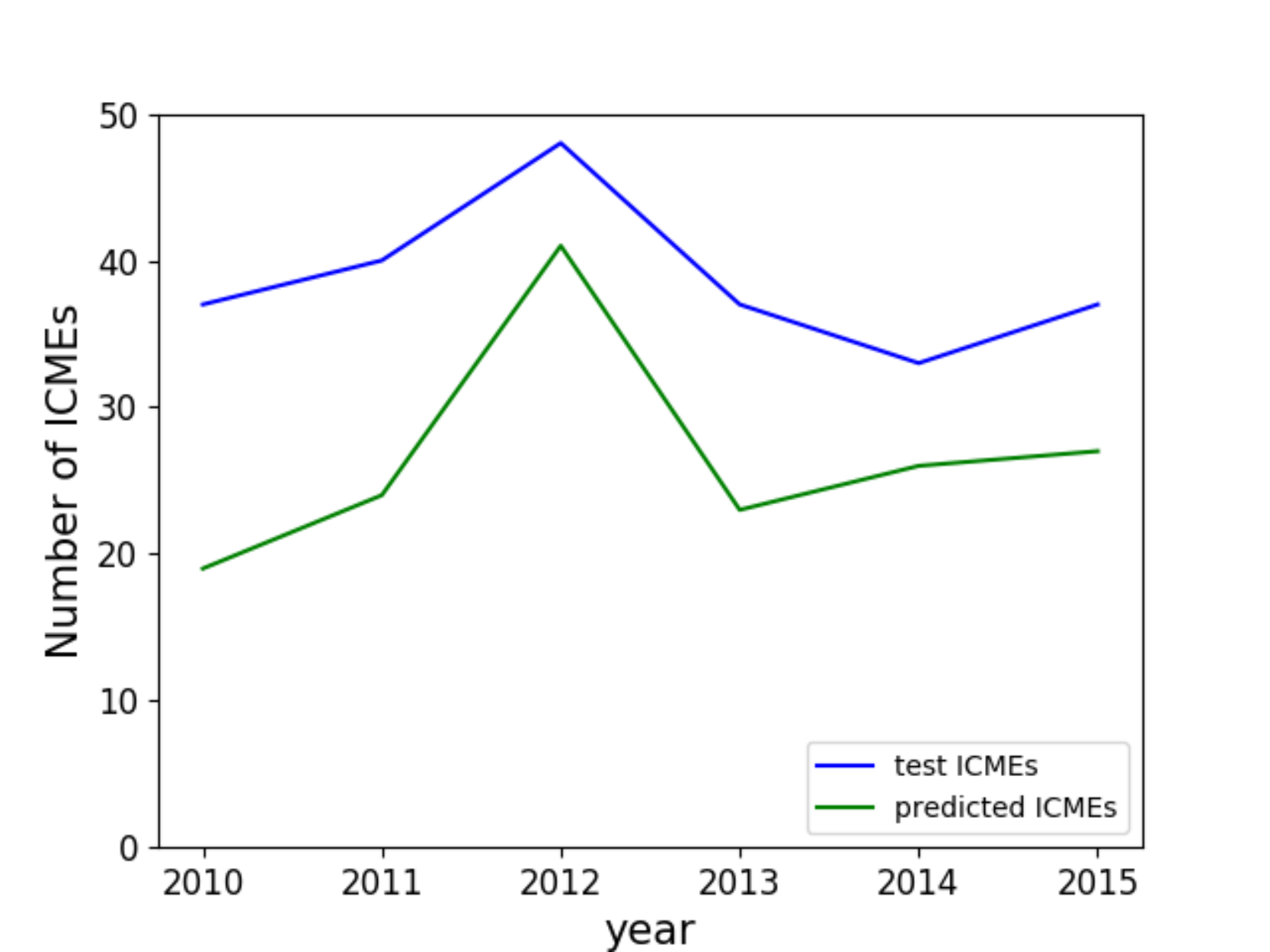}
\caption{Yearly occurrence frequencies of ICMEs of our catalog during our test period (blue) and the list predicted by our pipeline (green) for the high precision point on Figure \ref{precisionRecall}.}
\label{FPyearly}
\end{figure}

\begin{figure}
\centering
\includegraphics[scale=0.8]{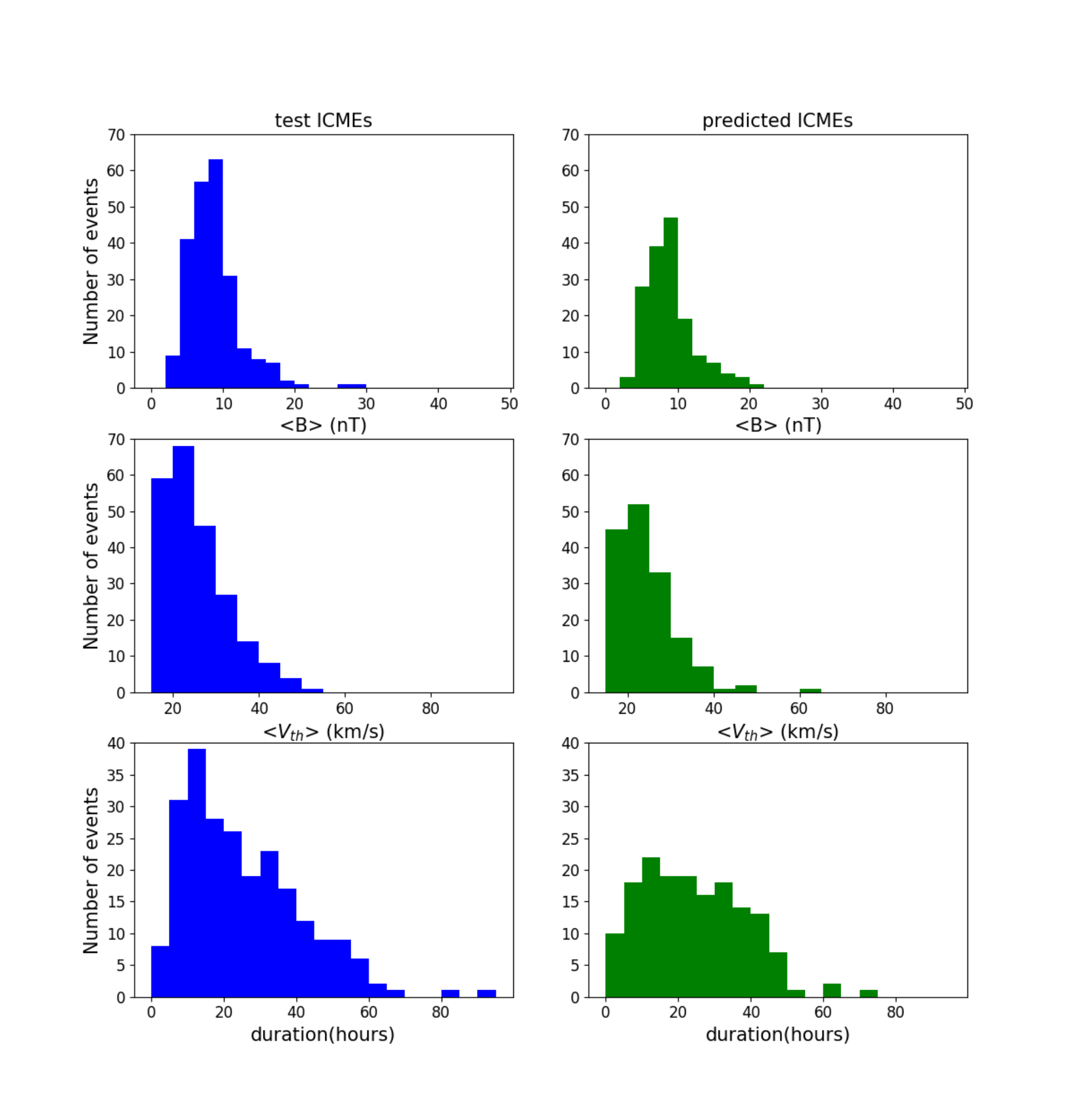}
\caption{Distribution of the mean values of different parameters of the ICMEs in our test period (blue) and of the predicted ICMEs in our high precision region(green). From top to bottom are represented: $\langle B\rangle$, $\langle V_{th}\rangle$ and the duration.}
\label{FPStats}
\end{figure}

The high precision region is represented by the yellow rectangle in the Figure \ref{precisionRecall}. In this region, the models generate an ICME list having a low number of FPs which ensures the consistency of the prediction. To quantify the performances of our method, we selected the working point shown in Figure \ref{precisionRecall} that corresponds to the decision threshold we used in the Figure \ref{GaussianFits}. At this point, 145 of the 232 ICMEs present in our test period are detected for a total recall of 62$\%$ and 25 of the 158 predicted ICMEs are considered as false positives for a total precision of 84$\%$. For this decision threshold, 120 of the 150 ICMEs present in the list of \citet{Chi16} were detected, which represents a recall of 80 $\%$ on this list. Likewise, we detected 82 of the 111 ICMEs present in the list of \citet{chinchilla} with a recall being equal to 74$\%$. It is then worth noting that even with a high value of precision we still manage to detect the great majority of the previously detected ICMEs. The 87 FNs we obtained in this case were distributed as follows:  21 of them were exclusive to the list of \citet{Chi16}, 20 came from \citet{chinchilla}, 9 were common to both list and 37 came from the ICMEs we discovered after different tests of our pipeline.
\\
Following the distribution of the FPs according the average reduced similarity in Figure \ref{histICMELikehighRecall}, all but one of the 25 FP we obtain, including the one represented in the Figure \ref{icmeLike}, will be considered as ICME-like and may contain one or several additional ICME we could add to our catalog.
\\
By increasing our decision threshold, we ensure our pipeline will return ICMEs that have been predicted with high values of similarities just as the one shown in Figure \ref{cloudRepresentation}. The generated predicted list is then supposed to contain easy-to-detect ICMEs that could be used for additional statistical study. To ensure it, we compared our predicted list to the ICMEs of our test period. Figure \ref{FPyearly} shows the yearly occurrence frequencies of the two catalogs. As expected, the predicted list is shorter than the test list because the increased value of the decision threshold reduced the number of predictions. The two lists follow the same trend and both of them peak at the solar maximum in 2012 which is a first argument for the consistency of our predicted list. Figure \ref{FPStats} represents the distribution of the mean values of the magnetic field $\langle B\rangle$, the thermal velocity $\langle V_{th}\rangle$ and the duration of our two catalogs, the left column corresponds to our test catalog while the right corresponds to our predicted list. Looking at the third row of subplots, the ICME we predict tend to be longer than the ICMEs of our catalog. This fact is partly due to the merge we did in our processing part as explained in the subsection \ref{subsection:automatization}. Nevertheless, the two first rows of supblots show similar distribution in magnetic field and thermal velocity. This confirms that our pipeline predicts consistent ICME that can be used by an external user for statistical studies.

\section{Robustness of the pipeline}
\subsection{Importance of the various features as ICME indicator}
To determine the relative importance of our different physical variables as important ICME indicators, we trained our pipeline on various configurations of our initial training set and compared the predictions that were made on our test set. The configurations we investigated are as follows : 
\begin{itemize}
\item By considering solely the magnetic field magnitude and components data
\item By considering the magnetic field data, the proton fluxes
 and the $\beta$
\item By considering the proton fluxes only
\item By considering the densities of protons and $\alpha$ particles only
\end{itemize}

\begin{figure}
\centering
\includegraphics[scale=0.75]{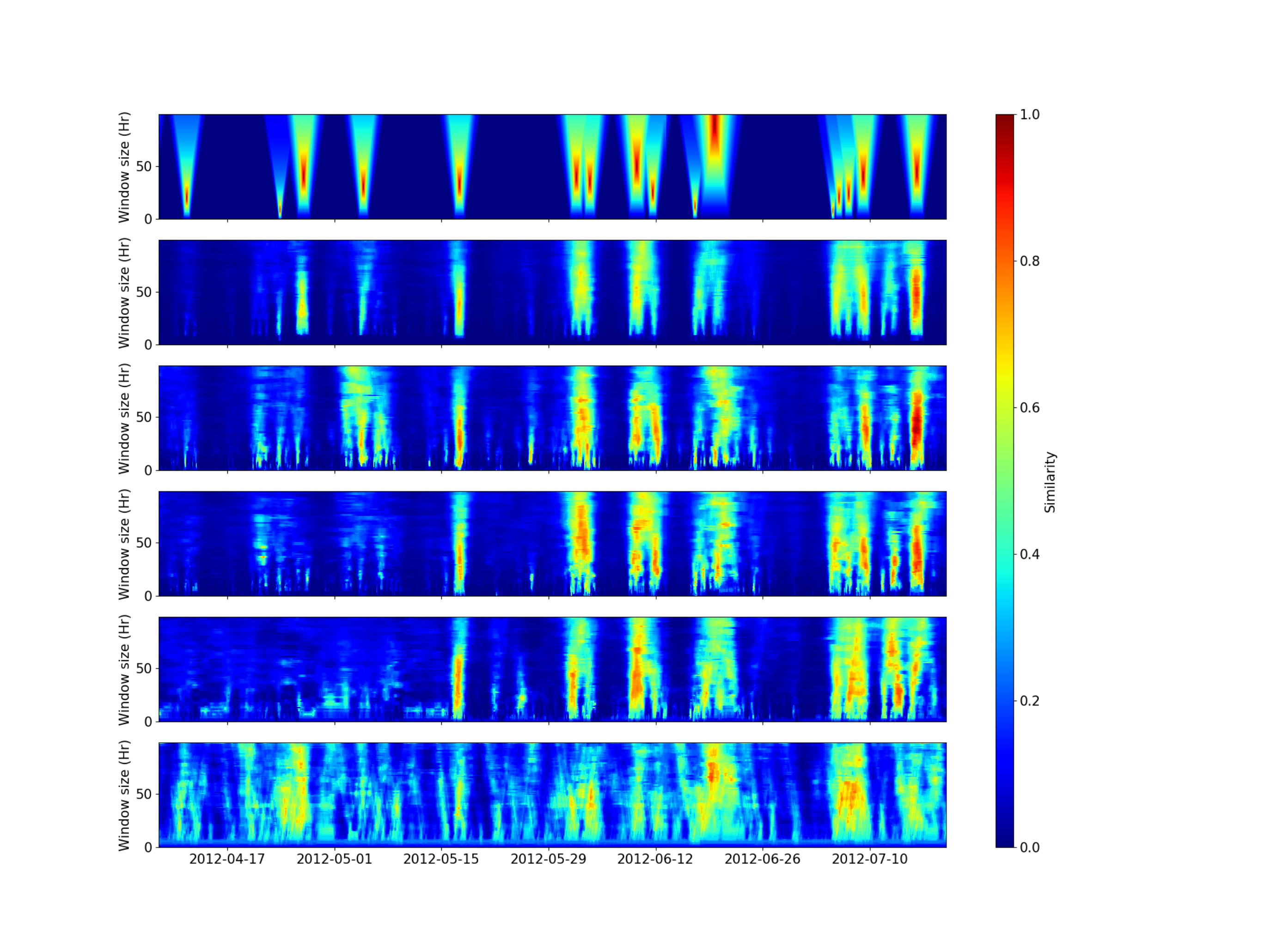}
\caption{Estimation of the similarity parameter for different sets of features during the period between the 3rd of July 2012 and the 17th of July 2012 From top to bottom are represented : the expected similarity, the prediction with the complete dataset, the prediction based only on the magnetic field and its components, the prediction based on the magnetic field, the plasma parameter $\beta$ and the proton fluxes, the prediction based on the lone proton fluxes and the prediction based on the lone measures of the protons and $\alpha$ particles.}
\label{featuresInfluence}
\end{figure}

Figure \ref{featuresInfluence} shows the similarity parameter which is expected (top panel) and predicted for the hundred windows from 1 to 100 hours in each of the above configurations on the same period between 8 April 2012 and 20 July 2012 as Figure \ref{predictedSimi}. On the right part of the figure, the ICMEs that had been predicted with the strongest values of similarity are still detected with high values of similarity for the three first different arrangements of features. This proves the ability of our method to detect an ICME with missing parameters which is to say when data from an instrument are not  available. Surprisingly, the lone measurements of densities can provide a fair detection of ICMEs despite the enhanced noise in this case. The prediction based on the densities values is even the only combination of features that detects the third ICME of Figure \ref{featuresInfluence} apart from the detection based on our complete dataset. Similarly, the prediction on the lone magnetic field components and amplitude is the only arrangement of features that detects the fourth ICME. As for visual detection of ICMEs, the magnetic field seems to play a key role in the CNN's learning. The possibility of detecting an ICME by using a specific set of features rather than an other is consistent with the possibility ICMEs have to partially fulfill the criteria generally used to detect them manually \citep{Zurbuchen06}. The fact that no ICME during the period of Figure \ref{featuresInfluence} detected by one of the subsets of features has not been also detected by our complete dataset indicates the importance each feature has in the characterization of the specific signatures of ICMEs as well as the importance of considering them altogether. 

\begin{figure}
\centering
\includegraphics[scale=0.6]{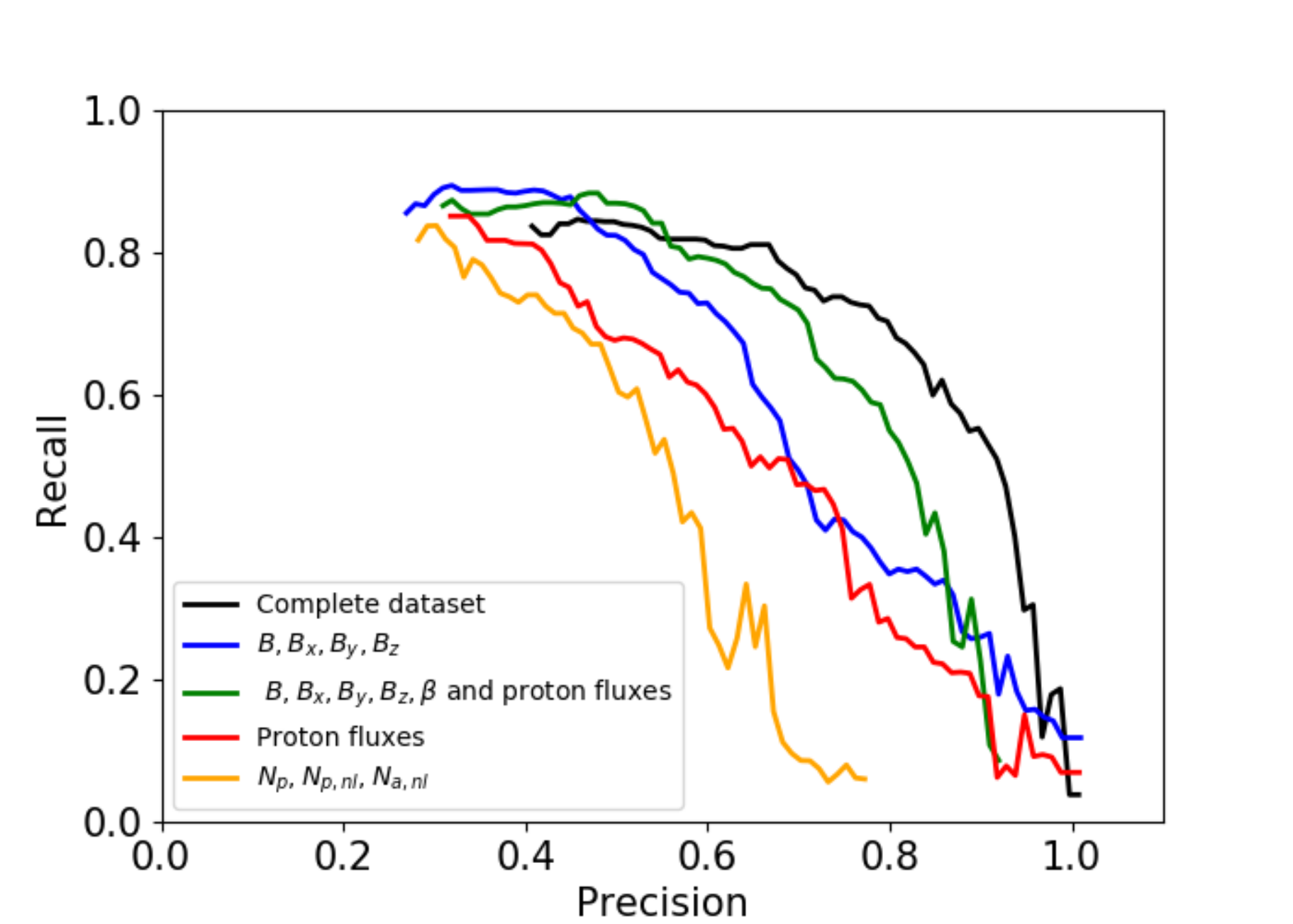}
\caption{Precision recall curves of our method using different dataset features : Our complete dataset (black), the lone magnetic field information (blue), the magnetic field, the plasma $\beta$ and the proton fluxes (green), the proton fluxes only (red) and the proton and $\alpha$ particles densities only (yellow).}
\label{featuresPrecisionRecall}
\end{figure}

To understand the impact of removing features, we compute the precision-recall curves for each configuration that are shown in Figure \ref{featuresPrecisionRecall}. The average precision is then computed for each dataset configuration, these values are shown in  Table \ref{featuresAUC}. Unsurprisingly, the highest value of this area is obtained for the complete dataset while the predictions  based on the proton fluxes and the one based on the densities have the lowest scores. This confirms the interest we have in considering the most complete set of features. The high values of the area obtained for the predictions based on the magnetic field, $\beta$ and the proton fluxes indicate the major importance these features have on the automatic detection of ICMEs.

\begin{table}
\centering

     \begin{tabular}{ |c| c|}
     \hline
  {\bf Dataset features} & {\bf Average precision} \\
  \hline
  All & 0.697 \\
  \hline
  $B, B_{x}, B_{y}, B_{z}$ & 0.593 \\
  \hline
  $B, B_{x}, B_{y}, B_{z}, \beta$ and proton fluxes & 0.621 \\
  \hline
  Proton fluxes only &  0.486\\
  \hline
  $N_{p}, N_{p,nl}, N_{a,nl} $ & 0.334\\
  \hline
	\end{tabular}
   \caption{Areas under the precision recall curve for different dataset configurations, higher values indicate more efficient detection.}
  \label{featuresAUC}
\end{table}

\subsection{Influence of the number of ICMEs in the training period}
\begin{figure}
\centering
\includegraphics[scale=0.6]{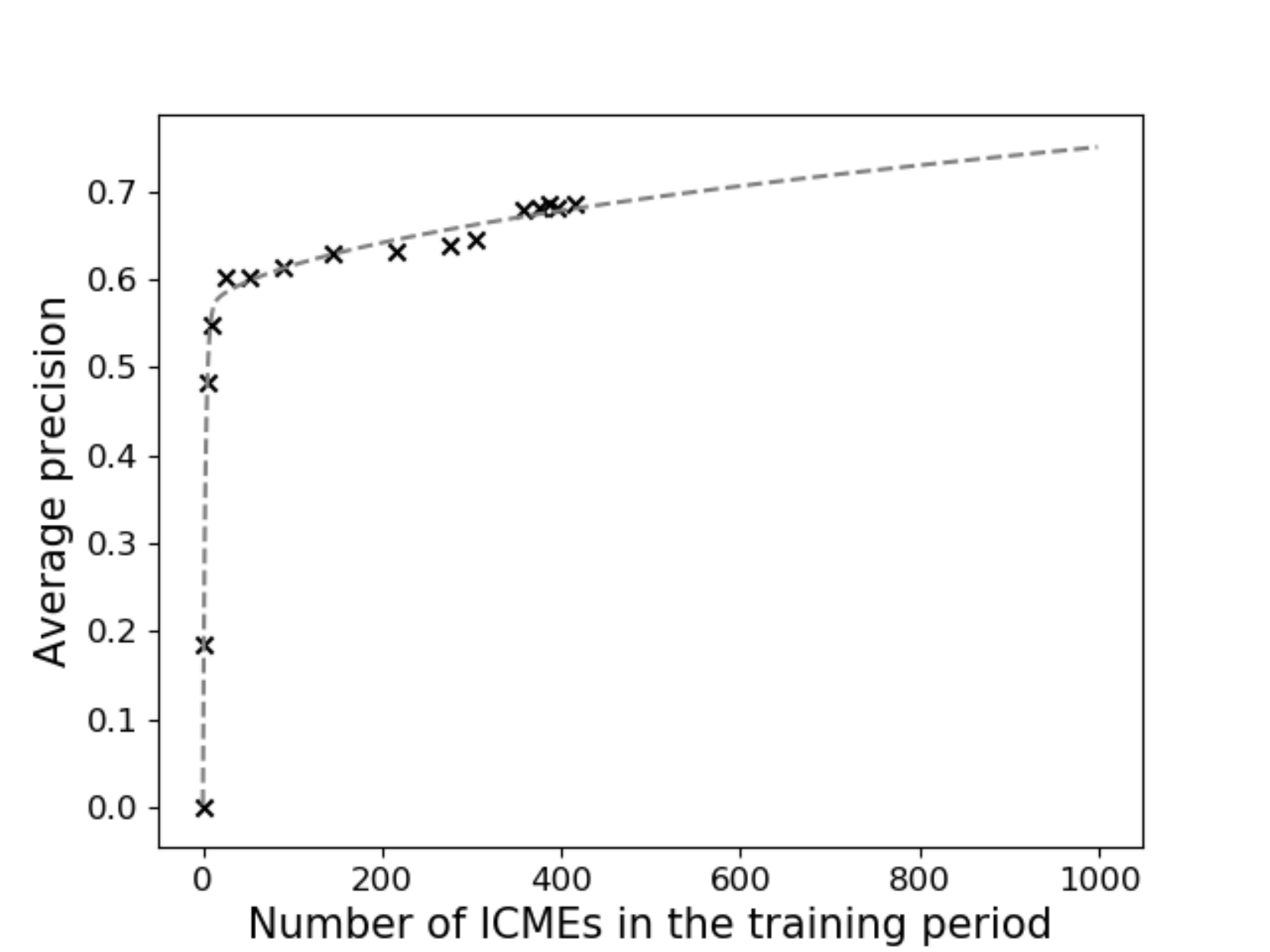}
\caption{Average precision of our pipeline as a function of the number of ICMEs present in the training period we considered, the tests we made are represented by the crosses. The grey dashed line represent the estimation of what the average precision would be if additional ICMEs were added to our training period.}
\label{fitPeriod}
\end{figure}

Statistical learning methods, in particular deep learning algorithms, often need a large quantity of data to give meaningful predictions. In this section, we investigate the influence of the number of ICMEs in our training period and its impact on the general performances of the method.
We thus progressively reduce the length of our training set by removing periods of data and thus ICMEs. This approach also allows us to investigate the real interest of considering a whole solar cycle of data during our training phase. Like in the previous subsection, we compute the average precision for each training period. The evolution of this value as a function of the number of ICMEs in the training period is shown in Figure \ref{fitPeriod}. The tests we made for various number of ICMEs are indicated with a cross x. The more we add ICMEs in our training period, the higher our average precision, which is consistent with the necessity of having a large quantity of data. However, the average precision starts with a sharp increase and evolves rapidly towards a weak inclination. The sharp increase indicates that a very low number of ICMEs are needed in order to reach fair performances. Surprisingly, it is even possible to start detecting ICMEs with the knowledge given by a single event as shown by the second point of the Figure \ref{fitPeriod}. It is very interesting to note that learning from few ICMEs with the complete set of features give similar performances than the whole training period with only particle densities or proton flux. Even if the progression is slower later-on, additional ICMEs keep improving the performances and it is then worth taking them into account.
From then on, we could estimate the number of ICMEs we would need in order to reach a certain of level of performances. This estimation is shown with the gray dashed line in the Figure \ref{fitPeriod}. At first sight, as many ICMEs as what we currently have would be needed for an increase in average precision by 10 $\%$. However, this expected number of ICMEs could be easily increased by completing our list with the FPs that appeared to be ICME-like and by extending our dataset period to the years before October 1997 and after 2015. Additionally, we showed in previous subsections the capacity our pipeline had to predict ICMEs with one or several missing features. Coupled with the diversity of spacecraft that have been providing in-situ measurements of ICMEs for the past 22 years (STEREO, 
Helios, Ulysse, ACE,...), one could perfectly imagine a dataset composed of the in-situ measurements provided by various spacecraft standardized in order to have consistent features and sampling for each spacecraft. This would increase drastically the number of given ICMEs and thus the performance of our pipeline. 

\subsection{Robustness regarding the training, the validation and the testing period}

As mentioned in the subsection \ref{subsection:recall}, we used our pipeline on three different training, validating and testing period. Changing the attribution of our three datasets period also has the interest of investigating the influence of the training period and the changes in WIND trajectory from 1997 to 2016 on our prediction. To do so, we changed the periods of our training, validation and test set as follows:

\begin{itemize}
    \item Training period from the 1st of April 2004 to the 31st of December 2015, validation between the 1st of January 2004 and the 30th of March 2004 and test on the 1997-2003 period
    \item Training on the 1997-2003 period and from from the 1st of April 2010 to the 31st of December 2015, validation between the 1st of January and the 30th of March 2010 and test on the 2004-2009 period
\end{itemize}

For each distribution, the average precision on the prediction made on the test set is shown on the Table \ref{periodAUC}. In the three cases we find similar values, which is consistent with the number of ICMEs (425, 530 and 353 respectively) contained in each training period and the average precision estimation provided by Figure \ref{fitPeriod}. This confirms the importance the diversity and the number of ICMEs seen by the CNNs during the training phase have on the quality of the prediction made by our pipeline.
Considering these three values, the mean average precision for our pipeline is then $0.694\pm 0.003$. 
Additionally, finding similar values for the three periods indicate that the changes in WIND trajectory especially during the 1997-2003 do not affect our results.
\begin{table}
\centering

     \begin{tabular}{ |c| c|}
     \hline
  {\bf Testing period} & {\bf Average precision} \\
  \hline
  2010-2015 & 0.697 \\
  \hline
  2004-2009 & 0.69 \\
  \hline
  1997-2003 & 0.694 \\
  \hline
	\end{tabular}
   \caption{Areas under the precision recall curve for different testing periods}
  \label{periodAUC}
\end{table}

\section{Global quality of the prediction}
The ambiguity that exists in the definition of the starting and ending times of an ICME combined of the non-exhaustivity of the different observers lists tends to limit the overview an event-based score would give on the quality of the detection made by our pipeline. It is thus interesting to also quantify to what extent the predicted list is globally similar to our list, and compare this global similarity to those of various independent expert list covering the same time period.

To do so, we define the Jaccard index between two lists $A$ and $B$ as 

\begin{equation}
    Jaccard(A, B) = \frac{duration(A \cap B)}{duration(A \cup B)}
\end{equation}

where the intersection and the union are defined as time intervals. With this definition, a low Jaccard  can be induced by both events exclusive to one of the two lists and the differences on the boundaries of the common events.

\begin{figure}
    \centering
    \includegraphics[scale=0.5]{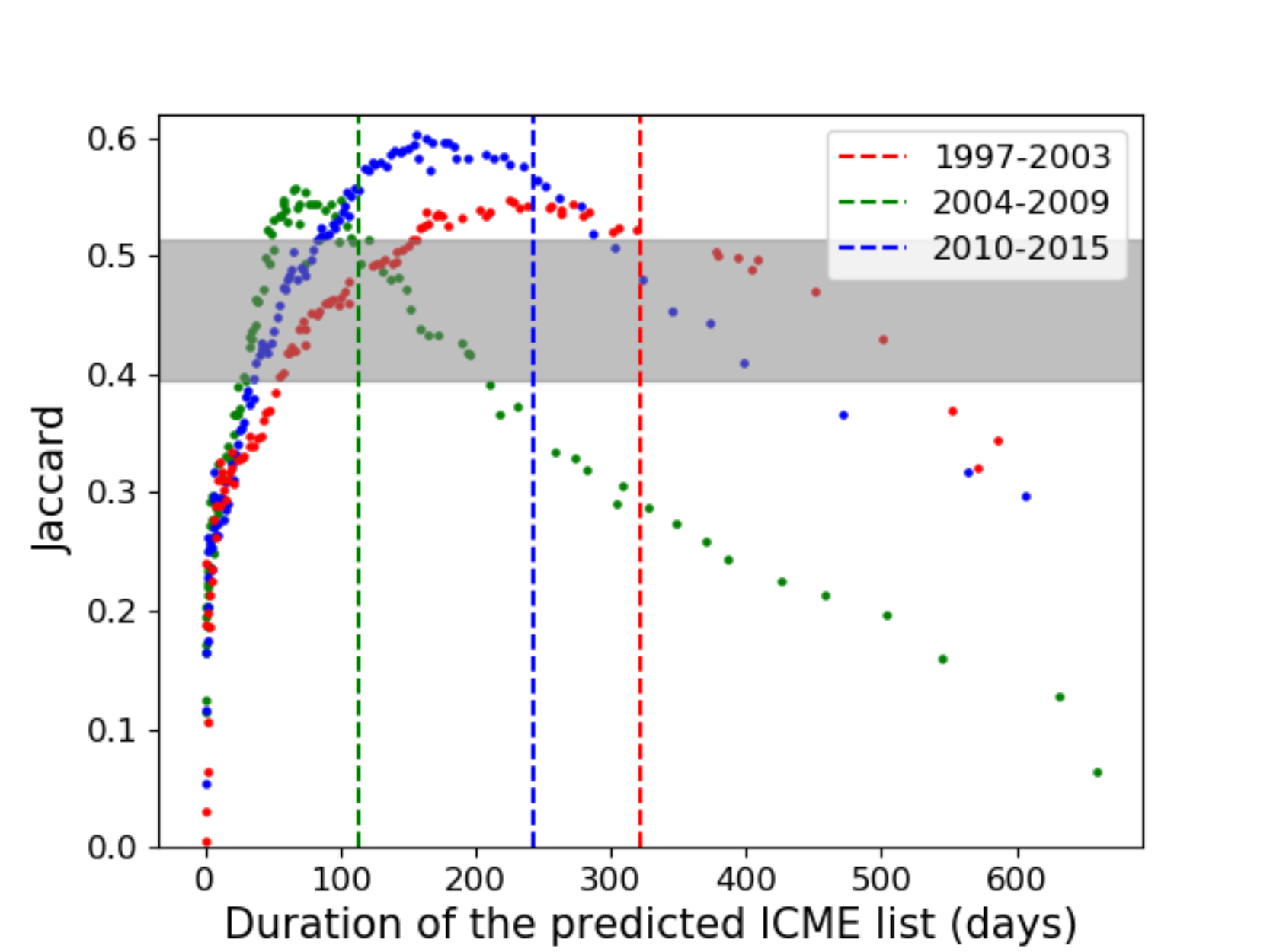}
    \caption{Jaccard between our ICME list and the generated ICME list as a function of the total duration of the generated ICME list (in days) for each prediction period : 1997-2003 (red), 2004-2009 (green) and 2010-2015 (blue). The vertical dashed lines represent the total duration of our list in each considered period. The gray line represents the confidence interval we have on the Jaccard between human made lists }
    \label{jaccard}
\end{figure}

For each prediction period, we computed the Jaccard index for each decision threshold on our reduced similarity and represented the evolution of this index as a function of the temporal size of the list predicted by our pipeline in Figure \ref{jaccard}.
High (resp. low) values of the total duration of the predicted list will correspond to a low (resp. high) decision threshold as the predicted list in this case will contain more (resp. less) events. The low value of the Jaccard in these cases is then mainly due to the important number of FPs (resp. FNs).
For each of our prediction period, we notice a similar evolution of the Jaccard that peaks around the temporal size of the reference list represented by the vertical dashed line (that will vary with the considered period according to Figure \ref{yearRepartLList}). This proximity can be understood as the peak will correspond to the best compromise we can find between a high recall and a high precision.
To compare the quality of our prediction regarding the global similarity that exists between different expert lists, we computed the Jaccard between lists that had the same number of events in one of the three prediction periods that we considered (e.g \citet{chinchilla, Chi16, RC10} and the RL for the three periods and \citet{Jian06} for the 1997-2003 and the 2004-2009 periods). The min-max interval for the different values of the Jaccard we found is represented on Figure \ref{jaccard} by the gray zone. The values we find that barely exceeds $50\%$ find their origin in the non-exhaustivity of the lists and the ambiguity that exists in the definition of the temporal boundaries of the ICMEs.
Below this interval, the generated list contains too much FPs or not enough events to give a significant insight about ICMEs.
Inside and above the interval, the list generated by our pipeline is as or more similar to our reference list than the human made lists between them. This is where the generated ICMEs lists can be used for further work such as the detection of additional events or statistical studies.

In the three cases, a non negligible part of the Jaccards are inside of above the typical expert list Jaccards lying in the gray zone. This proves that the lists generated by our pipeline are as globally similar to the RL as experts lists are to one another, and can then be used, either for further detections or statistical studies.

\section{Conclusion and perspective}
Using Convolutional Neural Networks that estimated a similarity parameter for windows of data of various sizes, from 1 to 100hr, and a post processing method based on peak detection, we developed a pipeline that provides an automatic ICME detection from the WIND spacecraft in-situ measurements. 
The 2D-similarity map the pipeline returned and that is shown in Figure \ref{predictedSimi} provides an interesting visual indicator of zones of interest for an external user particularly in the case of neighbored ICMEs or multiple events with various duration.

Our pipeline also has the ability to generate generic and reproducible ICME catalogs with a precision and a recall that has not been reached yet. From a Jaccard point of view, the list we predict are as comparable to our RL as two experts lists are together. Depending on the decision threshold we set on our detection, the pipeline offers the possibility to detect additional ICMEs (high-recall case) or to generate consistent and reproducible ICME catalogs that could be used for further statistical study (high-precision case).
From the insight we had on the FN in the high recall region, we showed that the ICMEs that are never detected by our pipeline are either short and too weak to be detected or  too close to another predicted ICME to be distinguished by the pipeline as a separate one.
Up to now, a total of 148 additional ICMEs have been detected  and were added to our WIND ICME catalog. Nevertheless, our catalog is not exhaustive yet and there are still ICMEs that have not been discovered yet. Additional runs of the pipeline shall be needed in order to establish a consistent ICME catalog as much exhaustive as possible.
\\
By testing our pipeline on datasets with missing features, we proved that even if the prediction was altered, our pipeline still has the capacity to detect ICMEs. On the first hand, this proves that our pipeline can be used even when one or several instruments of the spacecraft are defective in order to maintain a continuous prediction of ICMEs. On the other hand, this also proves that our pipeline can easily be adapted to the data of other spacecraft that have been measuring ICMEs over the past 22 years in different places of the solar system (Cluster, ACE, Stereo, Helios, Ulysse, Venus Express...) and for which measured features might be missing. 
\\
The influence of the number of ICMEs being present in the training set has been investigated by training our pipeline with reduced datasets. Even if a few number of ICMEs is enough to detect events properly, a large number of additional events is required if we want the quality of the prediction to improve significantly. These additional ICMEs could be added by considering additional training period such as the 2016-2018 period, by looking more precisely at the ICME-like FP that were found by our pipeline or even by considering the data provided by other spacecraft.
Another way we would have to improve our performances would stand in the fine tuning of the parameters of the CNN we used to make our prediction.
\\
Finally, the prediction of ICMEs has been established without giving to the algorithm any initial knowledge on ICME. The presence of an ICME in a given window of data being indicated to the algorithm through the similarity that only depends on the event temporal boundaries. This indicates the adaptability of our pipeline which could  be used to detect other phenomena likely to be measured by spacecrafts such as the sheaths of ICMEs, Co-rotating Interaction regions, Stream Interaction regions or even  signatures related to small scale processes such as magnetic reconnection.

\acknowledgments
\section*{Acknowledgements}
We thank the Paris-Saclay Center for Data Science for their technical support and the RAMP (\url{https://ramp.studio/}) held at INRIA Palaiseau in October 2018.
We also thank Vincent Genot, Philippe Garnier for the stimulating discussions. Finally, we thank the Centre français de Données sur la Physique des Plasmas spatiaux for the access to the visualization tools AMDA and 3D view.

%




 \appendix
 
 \section{Architecture of the Convolutional Neural Networks}
 
\begin{figure}
    \centering
    \includegraphics[scale=0.4]{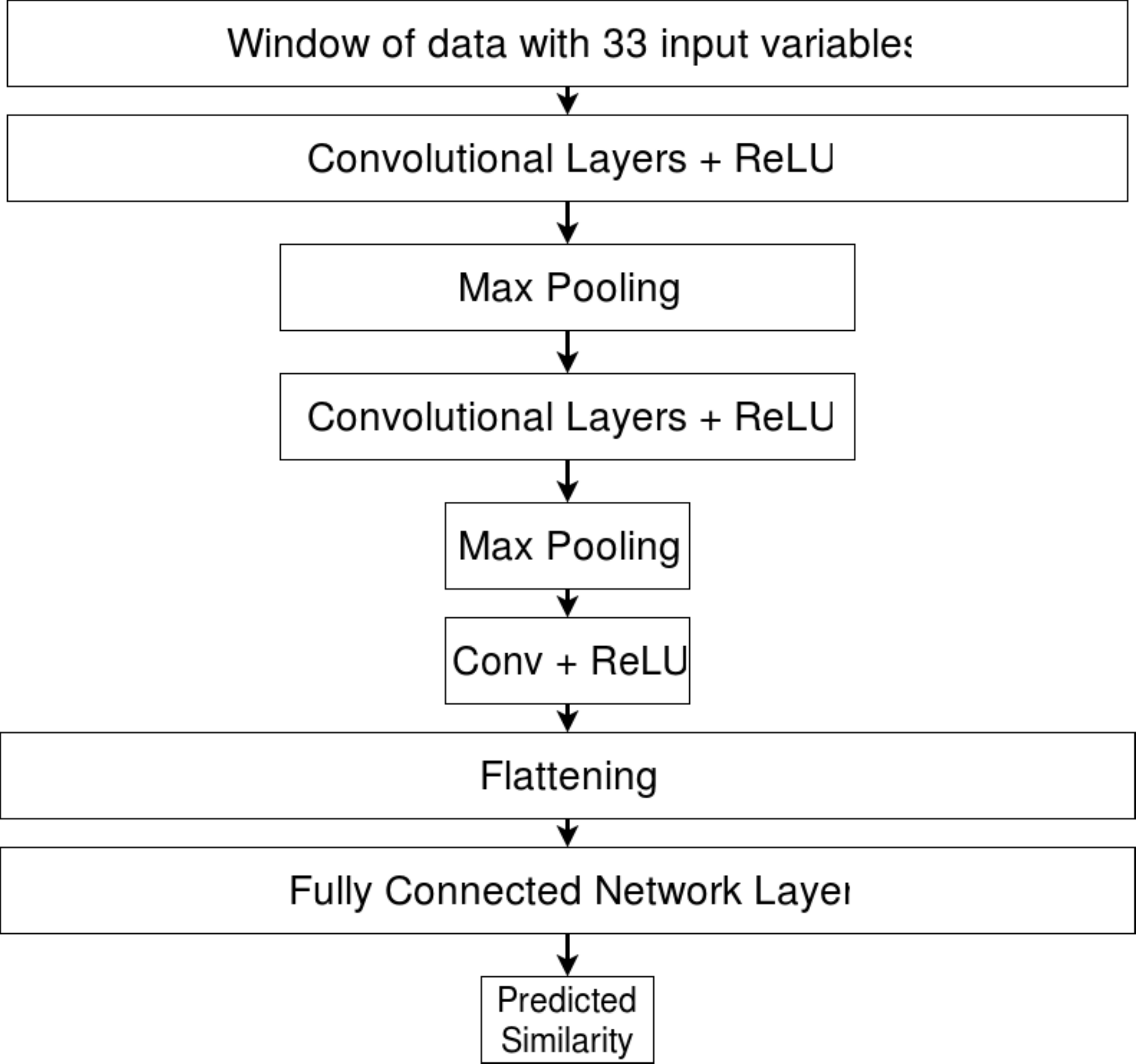}
    \caption{Illustration of the CNN architecture we used to estimate the similarity parameter for each window size}
    \label{cnn}
\end{figure}

 The typical architecture of the CNNs we used is shown in the Figure  \ref{cnn}. The different layers it contains can be grouped into four steps as detailed below. The first step consists of 1D convolutional layers which outputs are mapped by a Rectified Linear Unit (ReLU) layer. A max pooling layer then exhibits the maximum feature maps of the outputs this layer has the advantage of both reducing the dimension of the problem and reducing the overfitting the data. This operation is repeated twice. The second step is constituted by another convolutional and ReLu layer without Max Pooling. The third step  consists in unify the feature maps by concatenating them into an operation called Flattening. This output is then passed into a fully-connected network layer  that finally predicts the similarity of the window of data.
In order to avoid the overfit of data, a fraction of the nodes that constitute the fully-connected network layer are randomly dropped out at each step of the training phase. This operation ensures the robustness of the features learnt by the algorithm and, thus, reduce the overfit.

\bibliography{bibliography}



\end{document}